\documentclass[apj,twocolumn,floatfix]{openjournal}

\usepackage{xcolor}
\usepackage{textgreek}
\usepackage[utf8]{inputenc}
\usepackage[english]{babel}

\usepackage{hyperref}
\hypersetup{
    unicode, 
    colorlinks=true,
    linkcolor=linkcolor,
    citecolor=linkcolor,
    filecolor=linkcolor,
    urlcolor=linkcolor,
}
\usepackage{color,colortbl}
\definecolor{linkcolor}{rgb}{0.0,0.3,0.5}
\usepackage{tensind}
\tensordelimiter{?}
\DeclareGraphicsExtensions{.bmp,.png,.jpg,.pdf}
\usepackage{verbatim}
\usepackage[normalem]{ulem}
\usepackage{orcidlink}

\newcommand{\swift}{\textit{Swift }}

\usepackage{amsmath}

\graphicspath{{./}{figures/}}

\begin{document}

\title{A \textit{Swift} Fix II: Physical Parameters of Type I Superluminous Supernovae}


\author{\vspace{-1.3cm}Jason T. Hinkle\,\orcidlink{0000-0001-9668-2920}$^{1,*}$,
Benjamin J. Shappee\,\orcidlink{0000-0003-4631-1149}$^{1}$,
Michael A. Tucker\,\orcidlink{0000-0002-2471-8442}$^{2,3,4,\dagger}$
}

\affiliation{$^{1}$Institute for Astronomy, University of Hawai`i, 2680 Woodlawn Drive, Honolulu, HI 96822, USA\\
$^{2}$Center for Cosmology and Astroparticle Physics, The Ohio State University,  191 West Woodruff Ave, Columbus, OH, USA\\
$^{3}$Department of Astronomy, The Ohio State University,  140 West 18th Avenue, Columbus, OH, USA\\
$^{4}$Department of Physics, The Ohio State University, 191 West Woodruff Ave, Columbus, OH, USA\\
}

\altaffiltext{*}{FINESST Future Investigator}
\altaffiltext{$\dagger$}{CCAPP Fellow}

\email{jhinkle6@hawaii.edu}

\begin{abstract}

\noindent In November 2020, the \swift team announced a major update to the calibration of the UltraViolet and Optical Telescope (UVOT) data to correct for the gradual loss of sensitivity over time. Beginning in roughly 2015, the correction affected observations in the three near ultraviolet (UV) filters, reaching levels of up to 0.3 mag immediately prior to the correction. Over the same time period, an increasing number of Type I superluminous supernovae (SLSNe-I) were discovered and studied. Many SLSNe-I are hot (T$_\textrm{eff}$ $\approx 10,000$ K) near peak, and therefore accurate UV data are imperative towards properly understanding their physical properties and energetics. We recompute \swift UVOT photometry for SLSNe-I discovered between 2014 and 2021 with at least 5 \swift observations in 2015 or later. We calculate host-subtracted magnitudes for each SLSN and fit their spectral energy distributions with modified blackbodies to obtain the radius and temperature evolution. We also fit multi-band photometry using the Modular Open Source Fitter for Transients (MOSFiT) to obtain key parameters such as the spin period (P), magnetic field strength (B), ejecta mass (M$_\textrm{ej}$), and kinetic energy (E$_\textrm{kin}$). From our MOSFiT modeling, we also estimate the peak UV/optical luminosity (L$_\textrm{peak}$) and the radiative energy (E$_\textrm{rad}$). Under the assumption of magnetar-powered SLSNe, we find several strong trends, including anti-correlations between P and both L$_\textrm{peak}$ and E$_\textrm{rad}$, a correlation between E$_\textrm{kin}$ and E$_\textrm{rad}$, and an anti-correlation between B and E$_\textrm{rad}$.
\end{abstract}

\keywords{Core-collapse supernovae (304) --- Near ultraviolet astronomy(1094) --- Supernovae (1668)  --- Time domain astronomy (2109) --- Transient sources (1851)}

\section{Introduction} \label{sec:intro}

A core-collapse supernova (CCSN) marks the death of a massive star \citep[e.g.,][]{woosley02, heger03, smartt09}. The ``typical'' Type Ib/c and Type II CCSNe have been well-known for decades \citep[e.g.,][]{minkowski41, zwicky64, porter87}. However, a rare class of supernovae known as superluminous supernovae \citep[SLSNe; ][]{quimby07, quimby11, gal-yam12} has been observed over the past $\approx$15 years, with peak luminosities roughly $10-100$ times more luminous than normal Type Ia and core-collapse supernovae \citep[e.g.,][]{folatelli10}. The light curves of SLSNe often evolve more slowly than typical supernovae \citep{gal-yam19a}, on timescales of $\sim 20 - 80$ days \citep[e.g.,][]{nicholl17c, chen23a}.

Similar to normal SNe, the growing class of SLSNe can be divided into two main spectroscopic classes: those without hydrogen emission \citep[SLSN-I; ][]{quimby07, gal-yam12, nicholl17c, chen23a} and those with hydrogen emission \citep[SLSN-II; ][]{miller09, gezari09b}. Their superluminous nature notwithstanding, such events would otherwise be classified as SNe Ic and SNe IIn, respectively, in most cases. Nevertheless, SLSNe-I exhibit unique pre- and near-peak spectra with very blue continua and strong \ion{O}{2} absorption features \citep{quimby18, gal-yam19b}. Additional diversity in spectroscopic properties has been seen, such as the discovery of the SLSN-Ib subtype, lacking hydrogen lines but with strong helium absorption \citep{quimby18, yan20}.

The radioactive decay thought to power normal supernovae \citep[e.g.,][]{hoyle60, arnett82} cannot generally explain the luminosities of observed SLSNe \citep{quimby11}. As such, several more exotic models have been put forth. These include a central engine -- either the injection of energy from the spin-down of a magnetar \citep[e.g.,][]{kasen10, woosley10} or accretion onto a newly-formed black hole \citep{dexter13}, interactions between the SN ejecta and the circumstellar medium (CSM) \citep[e.g.,][]{chevalier11, moriya13}, and the radioactive decay of unusually large amounts of $^{56}$Ni from a pair-instability explosion \citep[e.g.,][]{barkat67, kasen11, woosley17}. As SLSN-II share many similarities with the Type IIn class of supernovae \citep{smartt09, gal-yam17}, they are most likely powered by interactions with abundant CSM \citep[e.g.,][]{inserra18}.

Conversely, the energy sources of SLSNe-I have proven more difficult to identify. Their spectra lack strong emission or absorption lines typically used to model SN photospheric evolution \citep{dessart16, woosley21} and many proposed theories predict similar observables \citep[e.g.,][]{sukhbold16}. Indeed, \citet{chen23b} find that a majority of SLSNe-I are equally well fit by magnetar spin-down and CSM$ + ^{56}$Ni decay models. A growing number of SLSNe-I exhibit bumps or undulations in their light curves, further confusing the problem. Such light curves are difficult to explain with a magnetar central engine \citep[e.g.,][]{nicholl17c}, although some efforts have been made to extend the magnetar model \citep{dong23}. Instead, the light curve undulations may be more naturally explained by unstable accretion onto a BH or CSM interactions, although spectral predictions for such models are lacking \citep[e.g.,][]{gal-yam19a}. Furthermore, the recently identified class of luminous supernovae \citep{gomez22}, with luminosities between those of typical CCSNe and SLSNe, can be powered by large amounts of $^{56}$Ni or weak magnetar engines, suggesting an underlying continuum.

Large samples of well-observed SLSNe are being compiled as all-sky surveys \citep[e.g., ASAS-SN, ATLAS, and ZTF][]{shappee14, tonry18, bellm19} and spectroscopic classification efforts \citep[e.g., PESSTO, SCAT][]{smartt15, tucker22} have expanded. This has allowed several population studies to be conducted \citep[e.g.,][]{nicholl17c, decia18, chen23b}, generally finding that magnetar models can describe the light curves of most SLSNe-I while finding considerable diversity in the population \citep[e.g.,][]{yan20, chen23a, chen23b}. Notably, many SLSNe only have observer-frame optical data, and those with observer-frame ultraviolet (UV) observations of SLSNe (often from \textit{Swift}) occurred during the period in which the UltraViolet and Optical Telescope (UVOT) sensitivity calibration was affected \citep[e.g.,][]{hinkle21b}. Given the strong UV emission of many SLSNe-I near peak, this motivates revisiting trends and correlations with corrected UV data.

The paper is organized as follows. In Section~\ref{sec:sample}, we discuss the sample selection, and in Section~\ref{sec:swift_reductions}, we discuss our reductions of the \swift UVOT data. In Section~\ref{sec:bb_fits} we describe our blackbody models of the SLSN SEDs. Section~\ref{sec:mosfit} details our modeling of the multi-band photometry with the Modular Open Source Fitter for Transients \citep[MOSFiT;][]{nicholl17c, guillochon18}. Section~\ref{sec:energy_corr} presents several correlations between physical parameters. Finally, we provide conclusions in Section~\ref{sec:conclusions}. Throughout this paper, we have used a cosmology with $H_0$ = 69.6 km s$^{-1}$ Mpc$^{-1}$, $\Omega_{M} = 0.29$, and $\Omega_{\Lambda} = 0.71$ \citep{wright06, bennett14}. 

\section{Sample Selection} \label{sec:sample}

\begin{deluxetable*}{cccccc}[h]
\tabletypesize{\footnotesize}
\tablecaption{Sample of Objects}
\tablehead{
\colhead{Object} &
\colhead{TNS ID} &
\colhead{Redshift} &
\colhead{Right Ascension} &
\colhead{Declination} &
\colhead{References} }
\startdata
ATLAS18unu$^a$ & SN2018ibb & 0.1586	& 04:38:56.950 & $-$20:39:44.10 & \citet{schulze24} \\
ATLAS18yff & SN2018hti & 0.063 & 03:40:53.76 & $+$11:46:37.38 & \citet{lin20} \\
ATLAS19ine & SN2019enz & 0.22 &	13:57:06.081 & $+$27:59:38.07 & \citet{19enz_epessto} \\
ATLAS19prf$^a$ & SN2019lsq & 0.14 & 00:04:40.6 & $+$42:52:11.35 & \citet{chen23a} \\
ATLAS19ynd$^a$ & SN2019szu & 0.213 & 00:10:13.14 & $-$19:41:32.46 & \citet{chen23a}, \citet{aamer24} \\
ATLAS20xqi$^a$ & SN2020rmv & 0.27 &	00:40:00.187 & $-$14:35:25.14 & \citet{chen23a} \\
ATLAS20zst & SN2020tcw & 0.0645	& 15:28:17.080 & $+$39:56:50.53 & \citet{20tcw_ZTF} \\
DES15S2nr & \dots & 0.22 & 02:40:44.62 & $-$00:53:26.4 & \citet{DES_atel} \\
Gaia16apd$^a$ & SN2016eay & 0.102 & 12:02:51.70	& $+$44:15:27.4 & \citet{nicholl17a, kangas17, yan17} \\
Gaia17biu & SN2017egm & 0.030721 & 10:19:05.620 & $+$46:27:14.08 & \citet{nicholl17b}, \citet{bose18a} \\
Gaia17cbp$^a$ & SN2017gci & 0.09 & 06:46:45.030	& $-$27:14:55.86 & \citet{fiore21} \\
Gaia18beg & SN2018bgv & 0.0795 & 11:02:30.290 & $+$55:35:55.79 & \citet{lunnan20} \\
iPTF15esb$^a$ & SN2016wi & 0.224 & 07:58:50.67 & $+$66:07:39.1 & \citet{liu17_slsn, yan17} \\
LSQ14mo & \dots & 0.253 & 10:22:41.53 & $-$16:55:14.4 & \citet{chen17} \\
OGLE16dmu & \dots & 0.426 &	04:48:26.34 & $-$62:20:10.6 & \citet{16dmu_atel} \\
PS15ae$^a$ & SN2015bn & 0.11 & 11:33:41.55	& $+$00:43:33.5 & \citet{nicholl16} \\
PS16aqv$^a$	& SN2016ard & 0.2025 & 14:10:44.558 & $-$10:09:35.42 & \citet{blanchard18} \\
PS16dnq	& SN2016els & 0.217	& 20:30:13.925 & $-$10:57:01.81 & \citet{16els_pessto} \\
PS22bca$^a$	& SN2021ahpl & 0.051 & 15:15:57.940 & $-$19:17:31.96 &  \citet{21ahpl_epessto} \\
ZTF19aawfbtg$^a$ & SN2019hge & 0.0866 &	22:24:21.21	& $+$24:47:17.12 & \citet{chen23a} \\
ZTF19abpbopt$^a$ & SN2019neq & 0.1075 & 17:54:26.736 & $+$47:15:40.62 & \citet{chen23a} \\
ZTF20abobpcb$^a$ & SN2020qlb & 0.159 & 19:07:49.550 & $+$62:57:49.61 & \citet{west23} \\
ZTF20acphdcg & SN2020znr & 0.1 & 07:19:06.420 & 23:53:07.37 & \citet{20znr_epessto} \\
ZTF20acpyldh & SN2020abjc & 0.219 &	09:28:00.274 & $+$14:07:16.62 & \citet{20abjc_blanchard} \\
ZTF21aaarmti & SN2021ek & 0.193	& 03:23:49.914 & $-$10:02:41.18 & \citet{21ek_epessto} \\
ZTF21abaiono & SN2021lwz & 0.065& 09:44:47.390 & $-$34:42:44.21 & \citet{2021lwz_tnsan} \\
ZTF21accwovq & SN2021zcl & 0.117 & 05:09:14.458	& $-$06:03:13.87 & \citet{21zcl_epessto} \\
\enddata 
\tablecomments{The 27 SLSNe-I we re-analyze in this manuscript. TNS ID is the ID given for objects reported on the Transient Name Server. References include the discovery papers and papers using \swift data taken in 2015 or later. For objects without a discovery paper or inclusion in a survey paper, we cite the initial classification of a SLSNe-I. \textit{If using the revised photometry presented here, please cite both this paper and the original paper(s) in which \swift photometry was published.}}
\tablenotetext{a}{SLSN with light curve undulations}
\label{tab:sample}
\end{deluxetable*}

To create our sample of supernovae, we searched both the Open Supernova Catalog \citep{Guillochon17} and the Transient Name Server\footnote{\url{https://www.wis-tns.org/}} (TNS) for spectroscopically-classified SLSNe-I discovered between 2014 and 2021, with \swift data taken in 2015 or later. As the \swift calibrations for data taken before 2015 are not significantly affected by the correction issued in November 2020, we do not consider any earlier SLSNe. We then limited our sample to objects that had five or more epochs of \swift UVOT photometry, to allow us to robustly measure the evolution of the UV fluxes. These UV epochs were generally observed near peak light. This threshold yielded 27 SLSNe-I. As its physical origin remains a matter of debate, we exclude the ambiguous source ASASSN-15lh \citep{dong16, leloudas16} from our sample. Table \ref{tab:sample} lists these objects, along with the appropriate references for the source classification and, when available, the discovery papers publishing \swift photometry. The redshifts for the SLSNe in our sample were typically taken from the Open Supernova Catalog or publicly available classification spectra on TNS, but for some sources without such spectra, we used the redshift listed in the appropriate discovery paper. The large majority of these redshifts were estimated from narrow host emission lines, with the remainder estimated from SN features. In Table \ref{tab:sample}, we also note which SLSNe have undulations in their light curves, either those noted in the literature or ones with clear undulations in survey light curves for sources without published papers.

\section{\swift UVOT Reductions}
\label{sec:swift_reductions}
Six of the \swift UVOT \citep{roming05, poole08} filters are typically used for photometric follow-up of transient sources: $V$ (5425.3 \AA), $B$ (4349.6 \AA), $U$ (3467.1 \AA), $UVW1$ (2580.8 \AA), $UVM2$ (2246.4 \AA), and $UVW2$ (2054.6 \AA). The wavelengths quoted here are the pivot wavelengths calculated by the SVO Filter Profile Service \citep{rodrigo12}, which we use throughout the remainder of this work. Many of the SLSNe in our sample have epochs with each of these filters, although some objects only used a subset of the full filter set. Additionally, for some objects, filters with early non-detections were dropped for late-time epochs.

The majority of UVOT epochs include multiple observations in each filter. We separately combined the images in each filter for each unique observation identification number using the HEASoft {\tt uvotimsum} package. We then used the {\tt uvotsource} package to extract source counts in a region centered on the position of the transient and background counts using a source-free region with a radius of $\approx 30-40''$. We used a source radius of $5''$ to minimize UVOT aperture corrections. We then converted the UVOT count rates into fluxes and magnitudes using typical calibrations \citep{poole08, breeveld10}. For each UVOT image, we confirmed that the source did not lie on a region of the detector with known sensitivity issues\footnote{\url{https://swift.gsfc.nasa.gov/analysis/uvot_digest/sss_check.html}} (also see the Appendix of \citealt[][]{edelson15}). Our raw \swift photometry, uncorrected for the host-galaxy flux contribution and Galactic foreground extinction, is shown in Table \ref{tab:raw_swift}.

\begin{deluxetable*}{cccccccc}[h]
\tablecaption{Unsubtracted \swift Photometry}
\tablehead{
\colhead{Object} &
\colhead{TNS ID} &
\colhead{MJD} &
\colhead{Filter} &
\colhead{Magnitude} &
\colhead{Uncertainty} &
\colhead{Flux Density} &
\colhead{Uncertainty}\\
&
&
&
&
& &
\colhead{(erg s$^{-1}$ cm$^{-2}$ \AA$^{-1}$)} &
\colhead{erg s$^{-1}$ cm$^{-2}$ \AA$^{-1}$)}}
\startdata
\dots & \dots & \dots  & \dots  & \dots & \dots & \dots  & \dots \\
ATLAS18unu & SN2018ibb & 58464.741 & V & 17.59 & 0.12 & 3.40E-16 & 3.74E-17 \\
ATLAS18unu & SN2018ibb & 58472.714 & V & 17.76 & 0.14 & 2.91E-16 & 3.73E-17 \\
\dots & \dots & \dots  & \dots  & \dots & \dots & \dots & \dots  \\
ATLAS18unu & SN2018ibb & 58464.737 & B & 17.76 & 0.08 & 4.52E-16 & 3.32E-17 \\
ATLAS18unu & SN2018ibb & 58472.711 & B & 17.82 & 0.08 & 4.28E-16 & 3.14E-17 \\
\dots & \dots & \dots  & \dots  & \dots & \dots & \dots  & \dots \\
ATLAS18unu & SN2018ibb & 58464.736 & U & 18.44 & 0.08 & 3.81E-16 & 2.79E-17 \\
ATLAS18unu & SN2018ibb & 58472.711 & U & 18.53 & 0.09 & 3.50E-16 & 2.89E-17 \\
\dots & \dots & \dots  & \dots  & \dots & \dots & \dots  & \dots \\
ATLAS18unu & SN2018ibb & 58464.733 & UVW1 & 19.78 & 0.11 & 2.00E-16 & 2.02E-17 \\
ATLAS18unu & SN2018ibb & 58472.710 & UVW1 & 20.14 & 0.13 & 1.43E-16 & 1.71E-17 \\
\dots & \dots & \dots  & \dots  & \dots & \dots & \dots  & \dots \\
ATLAS18unu & SN2018ibb & 58464.742 & UVM2 & 20.60 & 0.13 & 1.24E-16 & 1.48E-17 \\
ATLAS18unu & SN2018ibb & 58472.714 & UVM2 & 20.74 & 0.13 & 1.09E-16 & 1.30E-17 \\
\dots & \dots & \dots  & \dots  & \dots & \dots & \dots & \dots  \\
ATLAS18unu & SN2018ibb & 58464.737 & UVW2 & 20.86 & 0.14 & 1.17E-16 & 1.50E-17 \\
ATLAS18unu & SN2018ibb & 58472.712 & UVW2 & 21.29 & 0.19 & 7.85E-17 & 1.37E-17 \\
\dots & \dots & \dots  & \dots  & \dots & \dots & \dots  & \dots \\
\enddata 
\tablecomments{\swift photometry of the SNe without the host flux subtracted and with no correction for Galactic extinction. For epochs where the flux was less than a 3$\sigma$ detection, the magnitude column shows a 3$\sigma$ upper limit on the magnitude. All magnitudes are presented in the AB system, using published conversions for systems naturally in the Vega system. The data for each source are grouped by filter and sorted by increasing MJD. Here we show the SLSN ATLAS18unu (SN2018ibb) to illustrate the format. The full table is available as an ancillary file.}
\label{tab:raw_swift}
\end{deluxetable*}

\subsection{Host-Galaxy UV Contribution}

To compute accurate transient photometry, we require an estimate of the host-galaxy flux in each bandpass. By subtracting this host-galaxy flux from the \swift photometry, we can isolate the supernova flux. We estimated the host-galaxy flux in the \swift bands in two main ways. Some SLSNe-I had late-time \swift exposures of the host galaxy, often targeted specifically to estimate the host-galaxy flux. For these sources, we directly measured the \swift photometry of the host using the same source and background regions as the reductions for the supernova. These magnitudes are shown in Table \ref{tab:measured_swift_phot}.

\begin{deluxetable}{ccccc}[hb]
\tablecaption{Measured Host-Galaxy \swift Photometry}
\tablehead{
\colhead{Object} &
\colhead{TNS ID} &
\colhead{Filter} &
\colhead{AB Mag} &
\colhead{Uncertainty}}
\startdata
\dots & \dots & \dots  & \dots  & \dots \\
ATLAS18yff & SN2018hti & V & 17.79  & 0.17 \\
ATLAS18yff & SN2018hti & B & 19.88 & 0.45 \\
ATLAS18yff & SN2018hti & U  & 20.54 & 0.38 \\
ATLAS18yff & SN2018hti & UVW1 & 22.03 & 0.52 \\
ATLAS18yff & SN2018hti & UVM2 & 23.51 & 0.78 \\
ATLAS18yff & SN2018hti & UVW2 & 22.71 & 0.46 \\
\dots & \dots & \dots  & \dots  & \dots \\
\enddata 
\tablecomments{Measured photometry from \swift epochs without present transient flux. All magnitudes are presented in the AB system, using published conversions for \textit{Swift}. Here we show the SLSN ATLAS18yff (SN2018hti) to illustrate the format. The full table is available as an ancillary file.}
\label{tab:measured_swift_phot}
\end{deluxetable}

For SNe without late-time \swift data, we collected archival photometric data to fit the host-galaxy spectral energy distribution (SED) with stellar population synthesis models. We used gPhoton \citep{million16} to measure UV fluxes from Galaxy Evolution Explorer \citep[GALEX;][]{martin05} data. We obtained optical catalog photometry from the Sloan Digital Sky Survey (SDSS) Data Release 16 \citep[$ugriz$;][]{ahumada20} or Pan-STARRS \citep[$grizY$;][]{chambers16} depending on the source position. When possible, we obtained mid-infrared $W1$ and $W2$ magnitudes from the Wide-field Infrared Survey Explorer \citep[WISE;][]{wright10} AllWISE catalog. As the hosts of SLSNe-I are typically faint, dwarf galaxies at moderate redshift \citep[e.g.,][]{perley16, taggart21}, many hosts do not have solid photometry in all bands. Given the depths of the above UV/optical surveys, for sources where the host galaxy is undetected or only weakly detected in archival imaging, the host contribution to the measured flux in the \swift UVOT bands is negligible. We list the archival photometry used in the host-galaxy fits in Table \ref{tab:archival_phot}.

\begin{deluxetable}{ccccc}[htbp!]
\tablecaption{Archival Host-Galaxy Multi-wavelength Photometry}
\tablehead{
\colhead{Object} &
\colhead{TNS ID} &
\colhead{Filter} &
\colhead{AB Mag} &
\colhead{Uncertainty}}
\startdata
\dots & \dots & \dots  & \dots  & \dots \\
iPTF15esb  & SN2016wi  & NUV   & 22.84 &  0.24 \\
iPTF15esb  & SN2016wi  & g(SDSS) & 22.61 &  0.15 \\
iPTF15esb  & SN2016wi  & r(SDSS) & 21.90 & 0.15 \\
iPTF15esb  & SN2016wi  & i(SDSS) & 21.50 & 0.15 \\
iPTF15esb  & SN2016wi  & z(SDSS) & 21.44 & 0.54 \\
iPTF15esb  & SN2016wi  & W1    &  20.02  & 0.12 \\
iPTF15esb  & SN2016wi  & W2    &   20.77  & 0.50 \\
\dots & \dots & \dots  & \dots  & \dots \\
\enddata 
\tablecomments{Archival UV, optical, and infrared photometry used in the \textsc{FAST} SED fits for our objects. All magnitudes are presented in the AB system, using published conversions for systems naturally in the Vega system. Here we show the SLSN iPTF15esb (SN2016wi) to illustrate the format. The full table is available as an ancillary file.}
\label{tab:archival_phot}
\end{deluxetable}

We fit the available UV through IR photometry for each SN host with the Fitting and Assessment of Synthetic Templates code \citep[\textsc{FAST};][]{kriek09}. For our fits we assumed a \citet{cardelli89} extinction law with $\text{R}_{\text{V}} = 3.1$ and Galactic extinction at the coordinates of the host galaxy \citep{schlafly11}, a Salpeter IMF \citep{salpeter55}, an exponentially declining star-formation rate, and the \citet{bruzual03} stellar population models. To estimate the host-galaxy flux in each of the \swift UVOT filters, we computed synthetic photometry using the best-fit host SED from \textsc{FAST} and the UVOT filter response curves from the Spanish Virtual Observatory (SVO) Filter Profile Service \citep{rodrigo12}. To obtain uncertainties for the host-galaxy fluxes, we did Monte Carlo sampling by perturbing the archival host fluxes assuming Gaussian errors and running 1000 different \textsc{FAST} iterations for each host galaxy. The synthetic \swift UVOT magnitudes computed for each object are shown in Table \ref{tab:synth_phot}.

\begin{deluxetable}{cccccc}[h]
\tablecaption{Synthetic Host-Galaxy \swift Magnitudes}
\tablehead{
\colhead{Object} &
\colhead{TNS ID} &
\colhead{Filter} &
\colhead{AB Mag} &
\colhead{Uncertainty}}
\startdata
\dots & \dots & \dots  & \dots  & \dots  \\
iPTF15esb  &  SN2016wi  & V     & 22.09 & 0.11 \\
iPTF15esb  &  SN2016wi  & B    &  22.86 & 0.16 \\
iPTF15esb  &  SN2016wi  & U    & 23.19  & 0.22 \\
iPTF15esb  &  SN2016wi  & UVW1 & 23.22 & 0.25 \\
iPTF15esb  &  SN2016wi  & UVM2 & 23.27 & 0.25 \\
iPTF15esb  &  SN2016wi  & UVW2 & 23.41 & 0.25 \\
\dots & \dots & \dots  & \dots  & \dots & \\
\enddata 
\tablecomments{Synthetic host photometry computed from the Monte Carlo sampling of host galaxy SED fits with \textsc{FAST}. All magnitudes are presented in the AB system, using published conversions for \textit{Swift}. Here we show the SLSN iPTF15esb (SN2016wi) to illustrate the format. The full table is available as an ancillary file.}
\label{tab:synth_phot}
\end{deluxetable}

Two of our sources, SN2020rmv and SN2020abjc, have no optical survey detections of their host galaxy, consistent with their apparently hostless nature. This suggests that these SNe will not have considerable host-galaxy contamination in the \swift images. To confirm this, we used gPhoton \citep{million16} to compute 3$\sigma$ limits on the UV magnitudes at the SN location from pre-explosion GALEX data. For SN2020rmv, we find upper limits of $>$22.65 mag and $>$22.42 mag in the NUV and FUV bands, respectively. For SN2020abjc, the corresponding limits are $>$22.61 mag and $>$23.20 mag. Additionally, the host galaxy of SN2018ibb has an HST F336W upper-limit of $>$26.04 mag \citep{schulze24} and optical detections at the level of $\sim24$ mag. Given the limiting magnitude of \swift UVOT for a typical exposure time, the host galaxies for these sources contribute negligibly to the measured fluxes. Our \swift photometry with the host-galaxy flux contribution subtracted and corrected for Galactic foreground extinction is shown in Table \ref{tab:sutracted_swift}.

\begin{deluxetable*}{cccccccc}[htbp!]
\tablecaption{Host-Subtracted \swift Photometry}
\tablehead{
\colhead{Object} &
\colhead{TNS ID} &
\colhead{MJD} &
\colhead{Filter} &
\colhead{Magnitude} &
\colhead{Uncertainty} &
\colhead{Flux Density} &
\colhead{Uncertainty}\\
&
&
&
&
& &
\colhead{(erg s$^{-1}$ cm$^{-2}$ \AA$^{-1}$)} &
\colhead{erg s$^{-1}$ cm$^{-2}$ \AA$^{-1}$)}}
\startdata
\dots & \dots & \dots  & \dots  & \dots & \dots & \dots  & \dots \\
ATLAS18unu & SN2018ibb & 58464.741 & V & 17.50 & 0.12 & 3.62E-16 & 3.98E-17 \\
ATLAS18unu & SN2018ibb & 58472.714 & V & 17.67 & 0.14 & 3.09E-16 & 3.97E-17 \\
\dots & \dots & \dots  & \dots  & \dots & \dots & \dots  & \dots \\
ATLAS18unu & SN2018ibb & 58464.737 & B & 17.64 & 0.08 & 4.99E-16 & 3.66E-17 \\
ATLAS18unu & SN2018ibb & 58472.711 & B & 17.70 & 0.08 & 4.72E-16 & 3.46E-17 \\
\dots & \dots & \dots  & \dots  & \dots & \dots & \dots & \dots  \\ 
ATLAS18unu & SN2018ibb & 58464.736 & U & 18.30 & 0.08 & 4.33E-16 & 3.18E-17 \\
ATLAS18unu & SN2018ibb & 58472.711 & U & 18.39 & 0.09 & 3.99E-16 & 3.29E-17 \\
\dots & \dots & \dots  & \dots  & \dots & \dots & \dots & \dots  \\
ATLAS18unu & SN2018ibb & 58464.733 & UVW1 & 19.59 & 0.11 & 2.38E-16 & 2.41E-17 \\
ATLAS18unu & SN2018ibb & 58472.710 & UVW1 & 19.95 & 0.13 & 1.71E-16 & 2.04E-17 \\
\dots & \dots & \dots  & \dots  & \dots & \dots & \dots  & \dots \\ 
ATLAS18unu & SN2018ibb & 58464.742 & UVM2 & 20.33 & 0.13 & 1.59E-16 & 1.89E-17 \\
ATLAS18unu & SN2018ibb & 58472.714 & UVM2 & 20.47 & 0.13 & 1.39E-16 & 1.66E-17 \\
\dots & \dots & \dots  & \dots  & \dots & \dots & \dots  & \dots \\
ATLAS18unu & SN2018ibb & 58464.737 & UVW2 & 20.59 & 0.14 & 1.49E-16 & 1.91E-17 \\
ATLAS18unu & SN2018ibb & 58472.712 & UVW2 & 21.02 & 0.19 & 1.00E-16 & 1.75E-17 \\
\dots & \dots & \dots  & \dots  & \dots & \dots & \dots  & \dots \\
\enddata 
\tablecomments{\swift photometry of the transients with the host flux subtracted corrected for Galactic extinction. The uncertainties incorporate both the error on the photometry and from the host SED fits. For epochs where the transient flux was less than a 3$\sigma$ detection, the magnitude column shows a 3$\sigma$ upper limit on the transient magnitude. All magnitudes are presented in the AB system, using published conversions for systems naturally in the Vega system. The data for each source are grouped by filter and sorted by increasing MJD. Here we show the SLSN ATLAS18unu (SN2018ibb) to illustrate the format. The full table is available as an ancillary file.}
\label{tab:sutracted_swift}
\end{deluxetable*}

\section{Spectral Energy Distribution Fitting} \label{sec:bb_fits}

The SEDs of SLSNe show luminous UV/optical emission, with a UV excess well above what is typical for Type I supernovae, in some cases accounting for a majority of the emitted luminosity \citep[e.g.,][]{yan17}. In addition, this UV emission often persists at significant levels even after peak emission \citep[e.g.,][]{yan17, smith18}, extremely rare for less luminous classes of supernovae. As the optical spectra of SLSNe-I are relatively featureless, a blackbody is often a reasonable assumption for fitting the SED. This allows for straightforward estimates of the effective temperature and radius of the emitting region.

\subsection{Modified Blackbody Fits}

\begin{deluxetable*}{cccccccccccc}[htbp!]
\tablecaption{Modified Blackbody Fits}
\tablehead{
\colhead{Object} &
\colhead{TNS ID} &
\colhead{MJD} &
\colhead{log(L)} &
\colhead{dlog(L$_{l}$)} &
\colhead{dlog(L$_{u}$)} &
\colhead{log(R)} &
\colhead{dlog(R$_{l}$)} &
\colhead{dlog(R$_{u}$)} &
\colhead{log(T)} &
\colhead{dlog(T$_{l}$)} &
\colhead{dlog(T$_{u}$)}\\
\colhead{} &
\colhead{} &
\colhead{} &
\multicolumn{3}{c}{log([erg s$^{-1}$])} &
\multicolumn{3}{c}{log([cm])} &
\multicolumn{3}{c}{log([K])} 
}
\startdata
\dots & \dots & \dots  & \dots  & \dots & \dots & \dots & \dots & \dots & \dots & \dots & \dots \\
ATLAS18unu & SN2018ibb & 58464.7 & 44.293 & 0.026 & 0.025 & 15.822 & 0.036 & 0.035 & 3.949 & 0.012 & 0.012 \\
ATLAS18unu & SN2018ibb & 58472.7 & 44.265 & 0.030 & 0.029 & 15.853 & 0.038 & 0.038 & 3.926 & 0.012 & 0.013 \\
ATLAS18unu & SN2018ibb & 58476.4 & 44.269 & 0.033 & 0.031 & 15.866 & 0.041 & 0.040 & 3.921 & 0.013 & 0.014 \\
ATLAS18unu & SN2018ibb & 58481.0 & 44.300 & 0.041 & 0.039 & 15.950 & 0.050 & 0.050 & 3.887 & 0.017 & 0.016 \\
ATLAS18unu & SN2018ibb & 58484.7 & 44.234 & 0.034 & 0.035 & 15.894 & 0.043 & 0.046 & 3.898 & 0.015 & 0.015 \\
\dots & \dots & \dots  & \dots  & \dots & \dots & \dots & \dots & \dots & \dots & \dots & \dots \\
\enddata 
\tablecomments{Bolometric luminosity, effective radius, and temperature estimated from the modified blackbody fits to the host-subtracted and extinction-corrected \swift data. Here we show a subset of the fits for the SLSN ATLAS18unu (SN2018ibb) to illustrate the format. The full table is available as an ancillary file.}
\label{tab:BB_fits_tab}
\end{deluxetable*}

\begin{figure*}
\centering
 \includegraphics[width=0.33\textwidth]{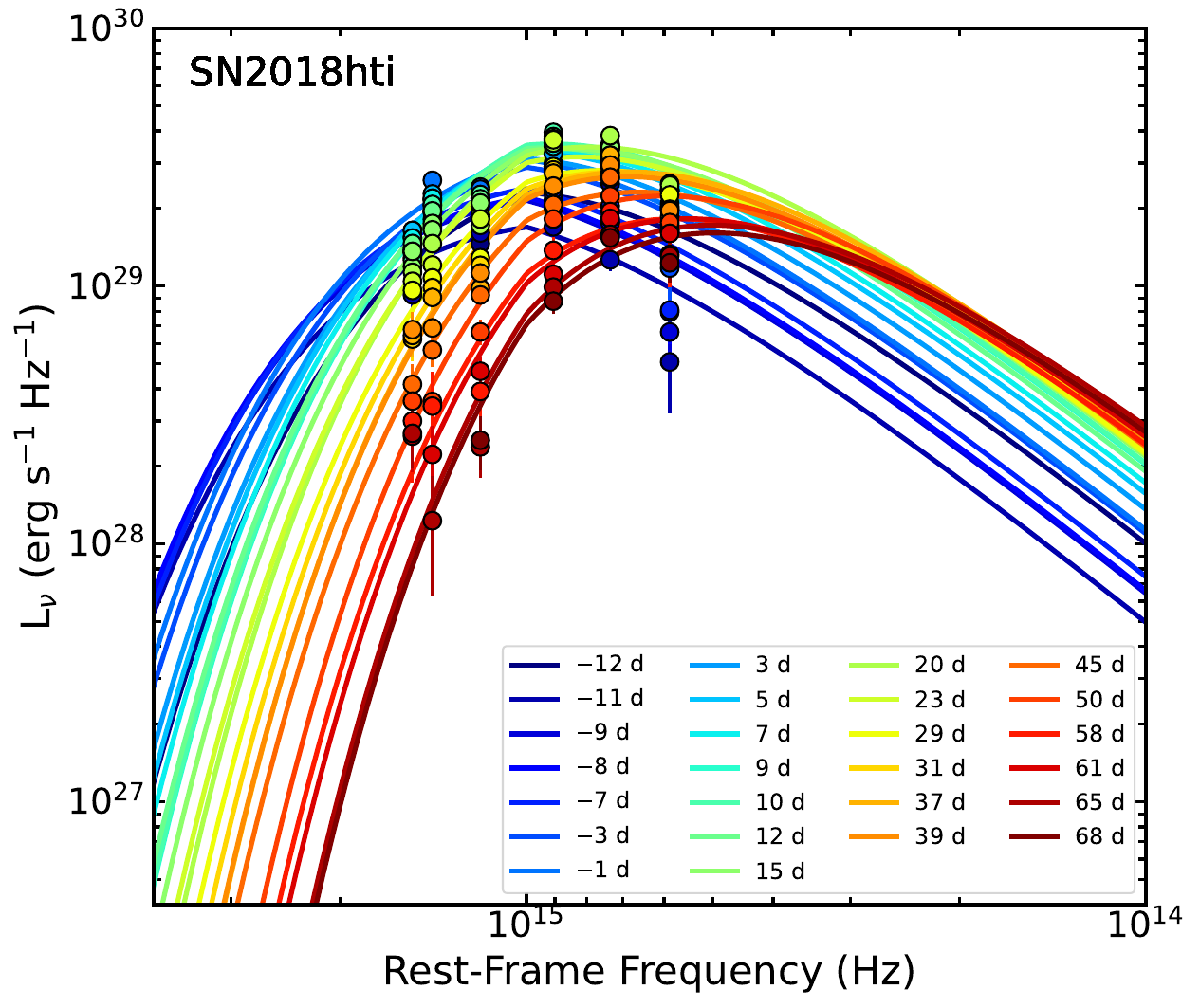}\hfill
 \includegraphics[width=0.33\textwidth]{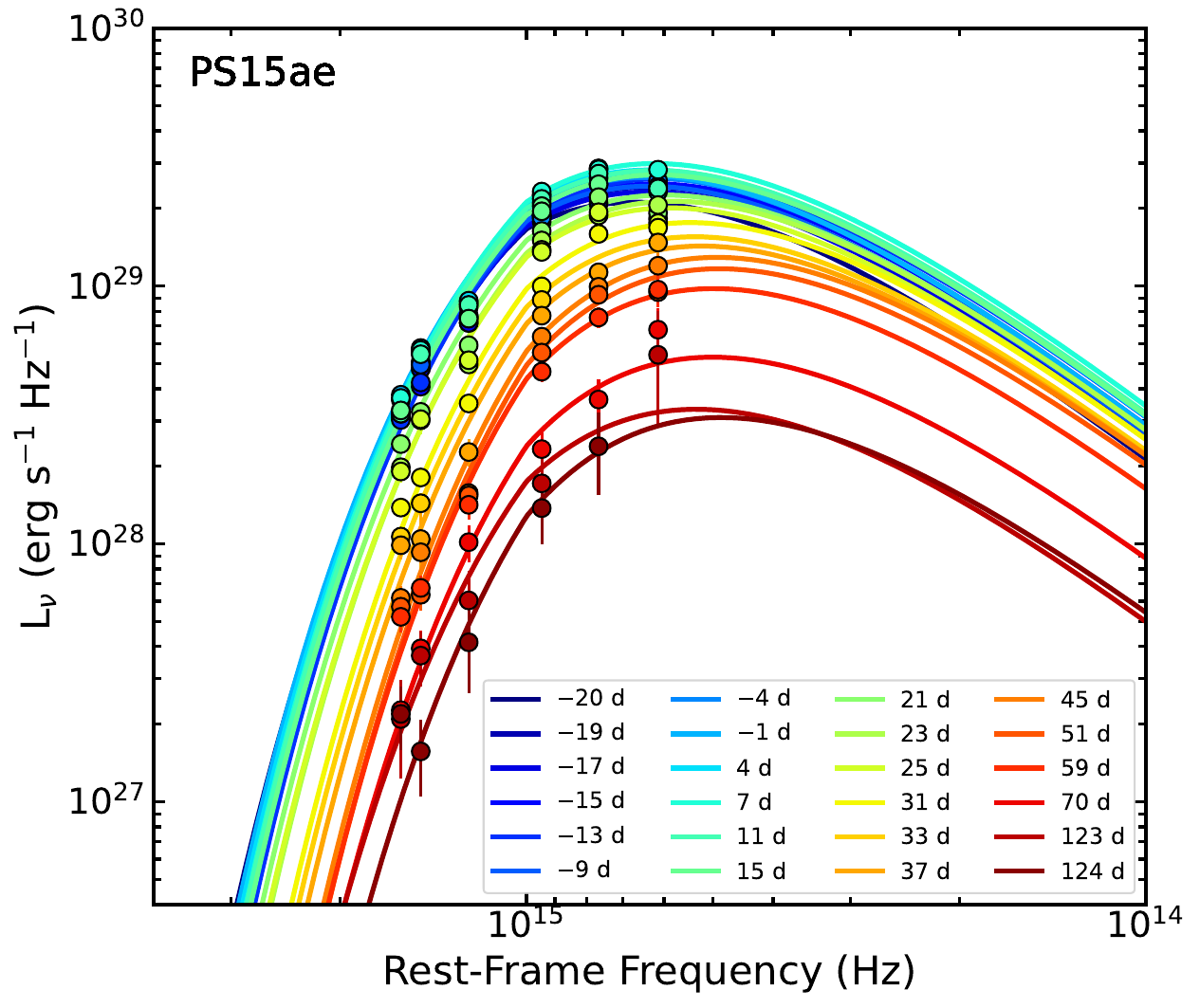}\hfill
 \includegraphics[width=0.33\textwidth]{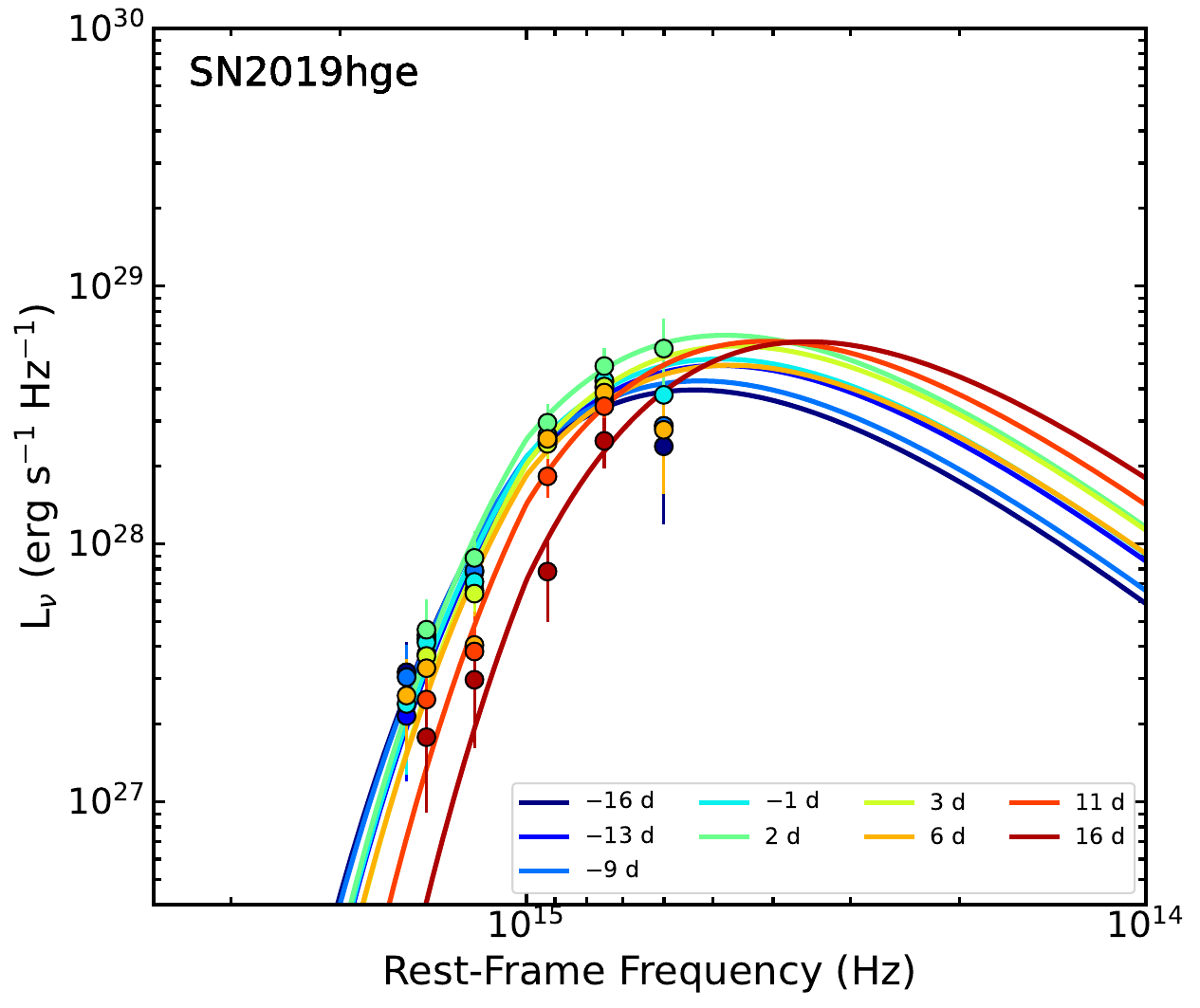}\hfill
 \caption{Evolution of the UV/optical photometry (points) and corresponding median modified blackbody fits (lines) for SN2018hti (left), PS15ae (middle), and SN2019hge (right). The times shown in the legends are rest-frame days relative to the UV/optical peak for each event.}
 \label{fig:fitted_BBs}
\end{figure*}

In the past several years, the number of rest-frame UV spectra of SLSNe-I has increased significantly, both due to deep surveys discovering faint objects at high redshift and all-sky surveys finding nearby, bright objects that can be observed by the Hubble Space Telescope \citep[e.g.,][]{yan18}. These UV spectra show two key features. One is a number of broad absorption features from species such as \ion{C}{2}, \ion{C}{3}, \ion{Ti}{3}, \ion{Si}{2}, and \ion{Mg}{2} \citep[e.g.,][]{yan17, yan18, smith18}. The other is that the FUV emission from SLSNe-I is suppressed as compared to a single blackbody fit to the optical and NUV emission \citep{chomiuk11, yan17, yan18}. This is likely due to a combination of blended absorption lines, metal-line blanketing \citep{hillier98, mazzali00}, and scattering of the UV photons within the expanding photosphere \citep{bufano09}. While the UV emission from SLSNe-I is suppressed, the line blanketing is significantly less than for SNe Ia, indicating low metallicity both in the progenitor star and the newly synthesized heavy element content of the ejecta \citep{yan17}.

Many studies fit the UV/optical SED using a modified blackbody to account for the suppression of the blackbody flux at short wavelengths. This ensures that blackbody fits to the full SED do not yield anomalously low temperatures. While a modified blackbody function with free parameters for the UV cutoff wavelength and slope \citep[e.g.,][]{yan18} can allow for statistically better fits, the increase in parameters relative to the small number of UV data points expands the uncertainties on the luminosity, temperature, and radius dramatically. Instead, we adopt the prescription of \citet{nicholl17c} to fit the \swift UVOT UV/optical SEDs of our supernovae. This form of the modified blackbody assumes a simple linear UV suppression at wavelengths below 3000 \AA, which is both a reasonable choice for typical SLSNe-I \citep{chomiuk11, yan17, yan18} and is consistent with the SED assumption for the light curve fitting with MOSFiT \citep{guillochon18, nicholl17c} to be discussed in Section \ref{sec:mosfit}.

We fit each epoch of \swift UVOT photometry using Markov Chain Monte Carlo (MCMC) methods \citep{foreman-mackey13} and a forward modeling approach. This accounts for the red leaks present in the $UVW2$ and $UVW1$ filters that may affect the photometry more significantly as the SN cools with time. We used the Spanish Virtual Observatories Filter Profile Service \citep{rodrigo12} to obtain the \swift UVOT filter response functions. We excluded ground-based optical data from our SED fits to both avoid de-weighting the UV data that is most important for an accurate temperature determination and mitigate cross-calibration issues between the \swift data and a heterogeneous sample of optical follow-up data. We only include data with $>2 \sigma$ detections to ensure robust luminosity and temperature estimates. As the SLSNe DES15S2nr, iPTF15esb, and ZTF21accwovq have no detected UV emission in their Swift epochs, we exclude them from the remainder of the analysis.

\begin{figure*}
\centering
 \includegraphics[width=0.96\textwidth]{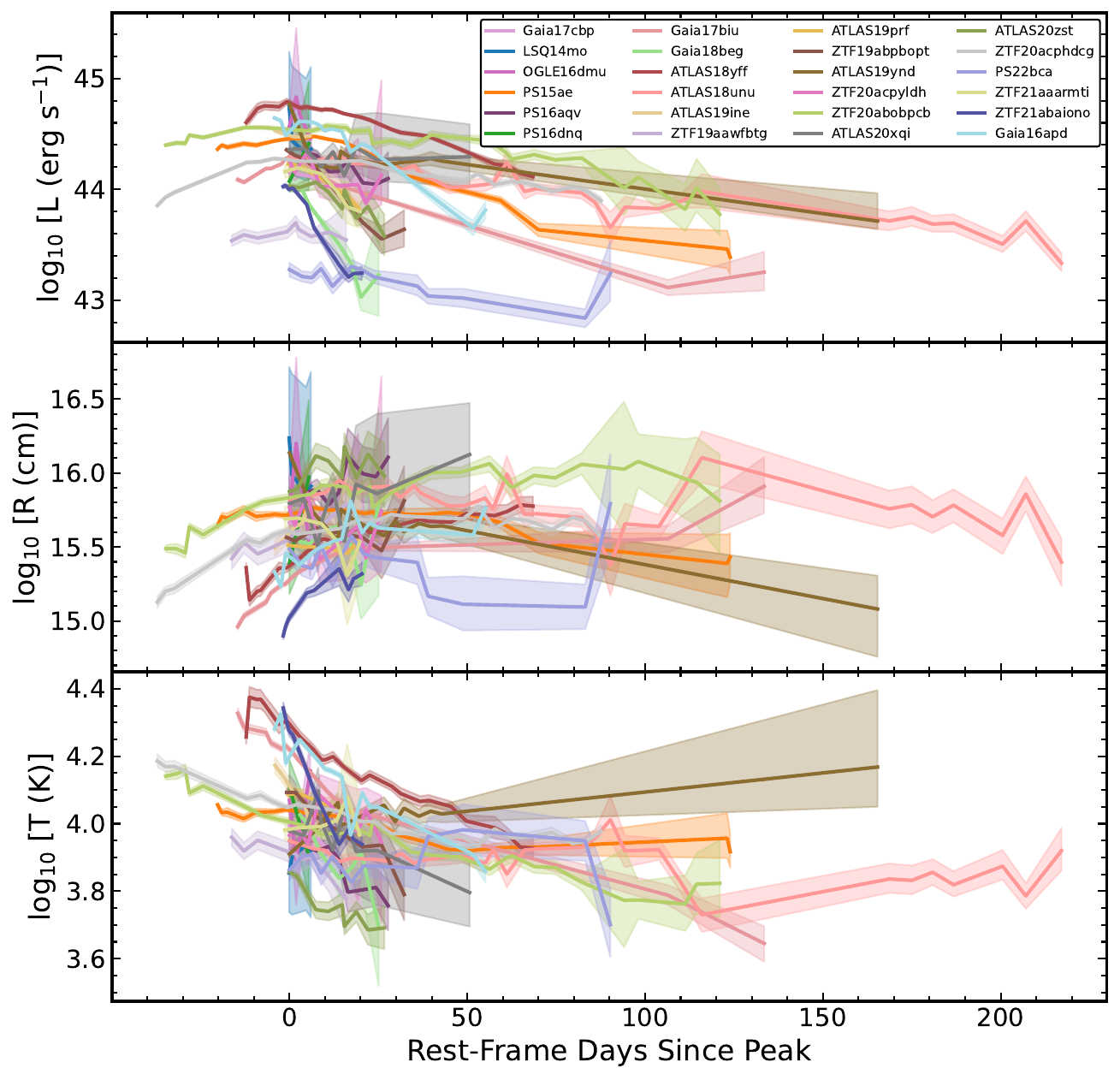}
 \caption{Temporal evolution of the UV/optical modified blackbody luminosity (top panel), radius (middle panel), and temperature (bottom panel) for the SLSNe-I in our sample. The solid lines are the median values, and the semi-transparent shading corresponds to the 1$\sigma$ uncertainty. The time is in rest-frame days relative to the time of peak luminosity.}
 \label{fig:BB_fits_fig}
\end{figure*}

The quality of the modified blackbody fits is illustrated in Figure \ref{fig:fitted_BBs}. The SEDs of SLSNe with high peak luminosities and rapid cooling, like SN2018hti, those with typical evolution, like PS15ae, and events with lower luminosities and fewer constraints, like SN2019hge, are all well fit by this model across a wide range of phases relative to peak. Figure \ref{fig:BB_fits_fig} shows the bolometric luminosity (L$_{\textrm{bol}}$), effective radius (R$_{\textrm{eff}}$), and effective temperature (T$_{\textrm{eff}}$) evolution for our sample of SLSNe, which are detailed in Table \ref{tab:BB_fits_tab}. We find a wide range of luminosities and decline rates within the luminosity evolution. Two outliers, one epoch each for OGLE16dmu and PS15ae, have not been shown in the figure, but are included in the table.

The blackbody radii generally increase with time as the SN ejecta expands. Some objects show a late-time plateau in R$_{\textrm{eff}}$, possibly indicating that they are entering the nebular phase \citep[e.g.,][]{nicholl19_slsn}. Conversely, the blackbody temperatures for nearly all objects decrease over time. The general temporal evolution seen in L$_{\textrm{bol}}$, R$_{\textrm{eff}}$, and T$_{\textrm{eff}}$ is consistent with blackbody fits to other samples of SLSNe-I \citep[e.g.,][]{lunnan18, chen23a}.

\subsection{Temperature and Radius at Peak Light}

\begin{figure*}
\centering
 \includegraphics[height=0.48\textwidth]{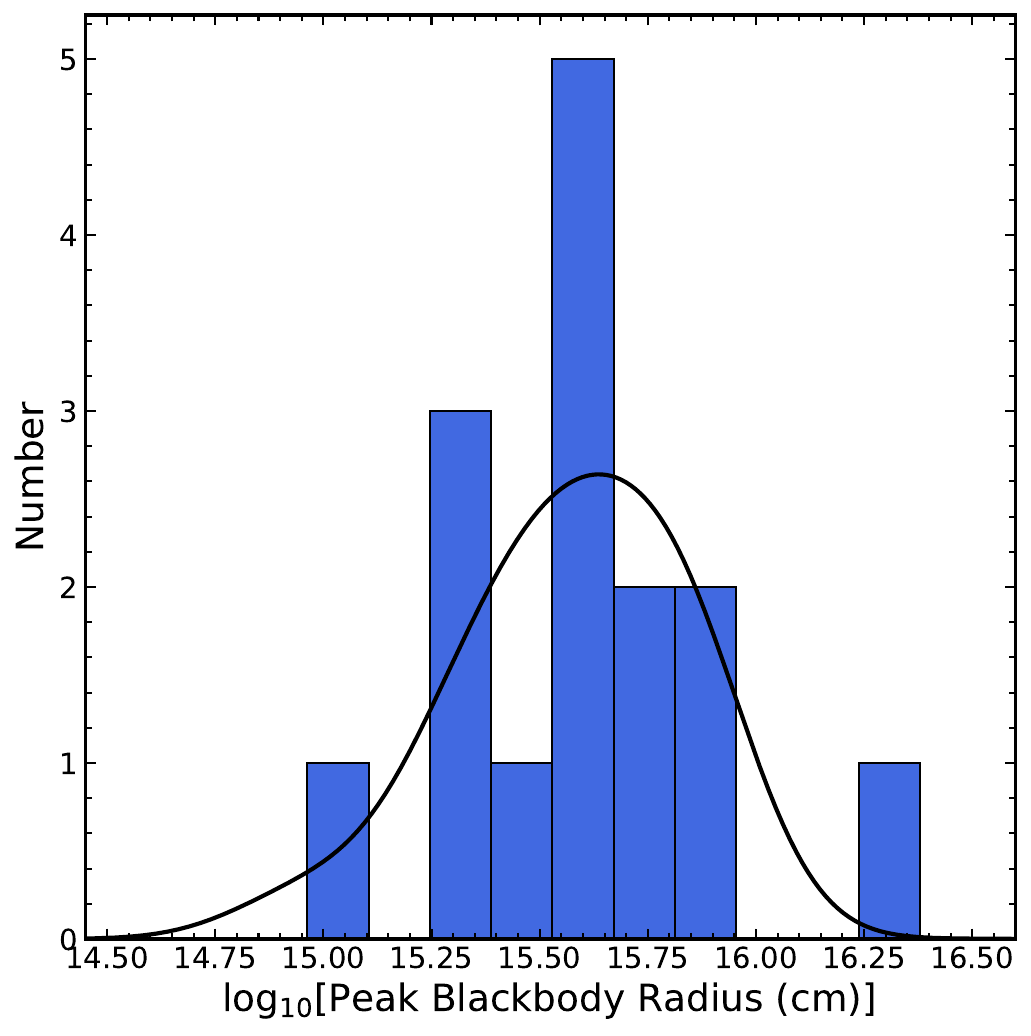}
 \includegraphics[height=0.48\textwidth]{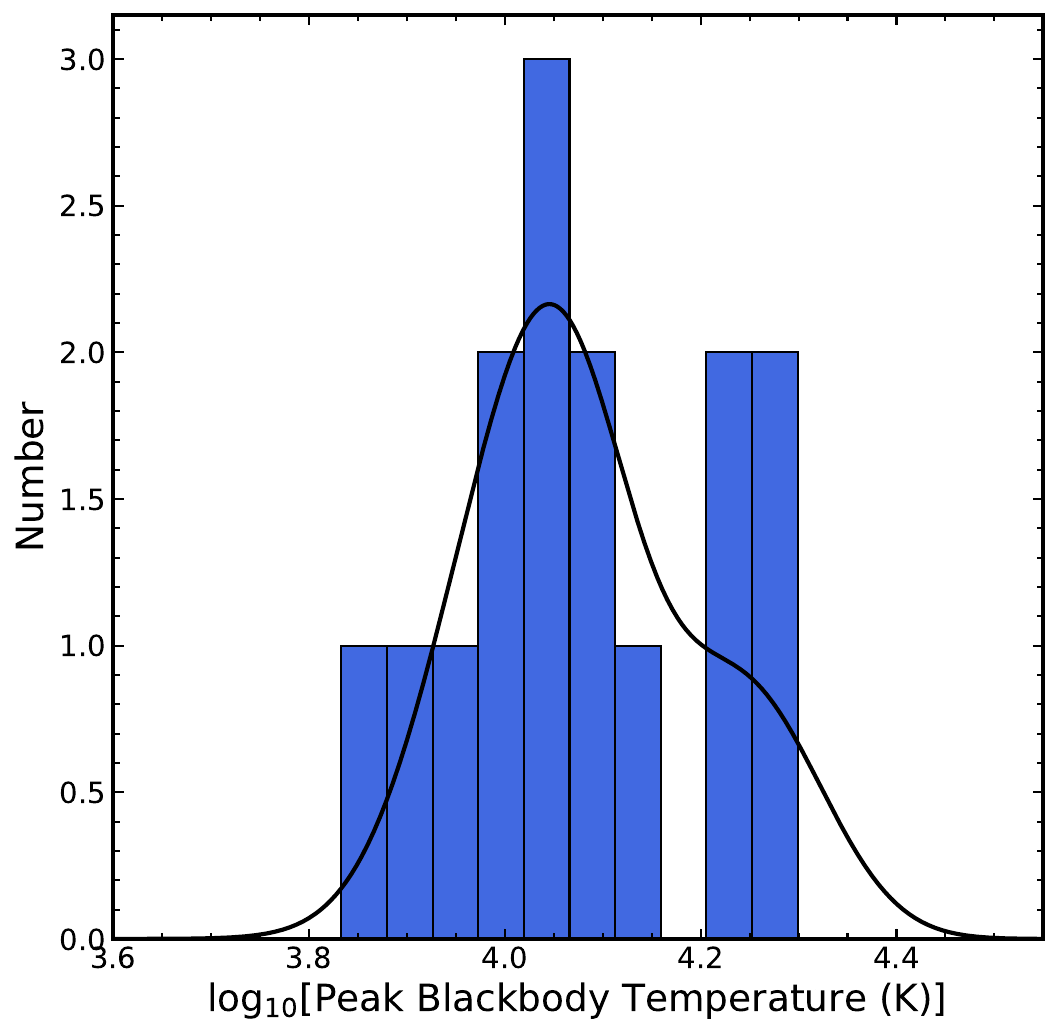}
 \caption{Histograms of R$_{\textrm{eff}}$ and T$_{\textrm{eff}}$ at the time of peak luminosity. Shown in black are KDEs of each distribution normalized to the sample size. The individual SN contribution to the KDEs is weighted by the inverse square of the estimated uncertainty on the peak radius and temperature, with a 1\% error floor added in quadrature to avoid over-weighting single objects.}
 \label{fig:RT_hist}
\end{figure*}

With the temporal evolution of the modified blackbody parameters shown in Figure \ref{fig:BB_fits_fig}, we calculated the temperature and radius at the peak bolometric emission for each SN. To do this, we first bolometrically corrected optical ground-based light curves for each supernova by scaling the optical photometry to match the interpolated bolometric luminosity derived from the modified blackbody fits, similar to previous transient studies \citep[e.g.,][]{holoien20, hinkle21a}. From this higher cadence bolometric light curve, often with pre-peak constraints, we fit for the time of peak luminosity using a generic magnetar model \citep{ostriker71, kasen10}. We first fit the full light curve to establish an initial estimate of the peak time and then restricted the fit to within $-$20 and $+$30 days of the estimated peak. We found this to generally return reliable estimates of the peak time. For two objects, SN2016els and SN2021ahpl, the magnetar model fits did not converge, and we instead took the epoch of maximum luminosity as our estimate of the peak time.

We then measured the temperature and radius at peak. We linearly interpolated the temperature with 5 days of peak on either side and then fit a line. The temperature and radius at peak were taken to be the value of this line at the time of peak emission. We estimated the uncertainty on these values by taking the standard deviation of 3000 Monte Carlo iterations of this linear fit. To ensure robust estimates of the temperatures and radii at peak, we only show the 13 supernovae with \swift data prior to peak and two objects, PS16aqv and SN2018bgv, having \swift data within 5 rest-frame days of peak. The rest of the sample either has the first \swift epoch more than 5 rest-frame days after peak or has a highly uncertain time of peak.

Histograms of T$_{\textrm{eff}}$ and R$_{\textrm{eff}}$ at peak are shown in Figure \ref{fig:RT_hist}, excluding the SLSNe without \swift data sufficiently close to peak light or poor constraints on the time of peak. Along with the histograms, we show kernel density estimates (KDE) of the underlying distribution computed using \textsc{scipy.stats.gaussian\_kde} and Scott's Rule \citep{scott92}. We find that R$_{\textrm{eff}}$ at peak spans $(9-200)\times 10^{14}$ cm with a peak in the radius distribution at $\approx 4 \times 10^{15}$ cm, consistent with previous results \citep[e.g.,][]{lunnan18, chen23a}. T$_{\textrm{eff}}$ values at peak span $\approx 7-20$ kK with a central peak at $\approx 11$ kK and a hotter inflection in the distribution at $\approx 18$ kK, all consistent with earlier work \citep[e.g.,][]{lunnan18, chen23a}.

\subsection{Comparison of Blackbody and Modified Blackbody Fits}

\begin{figure}[ht]
\centering
 \includegraphics[width=0.49\textwidth]{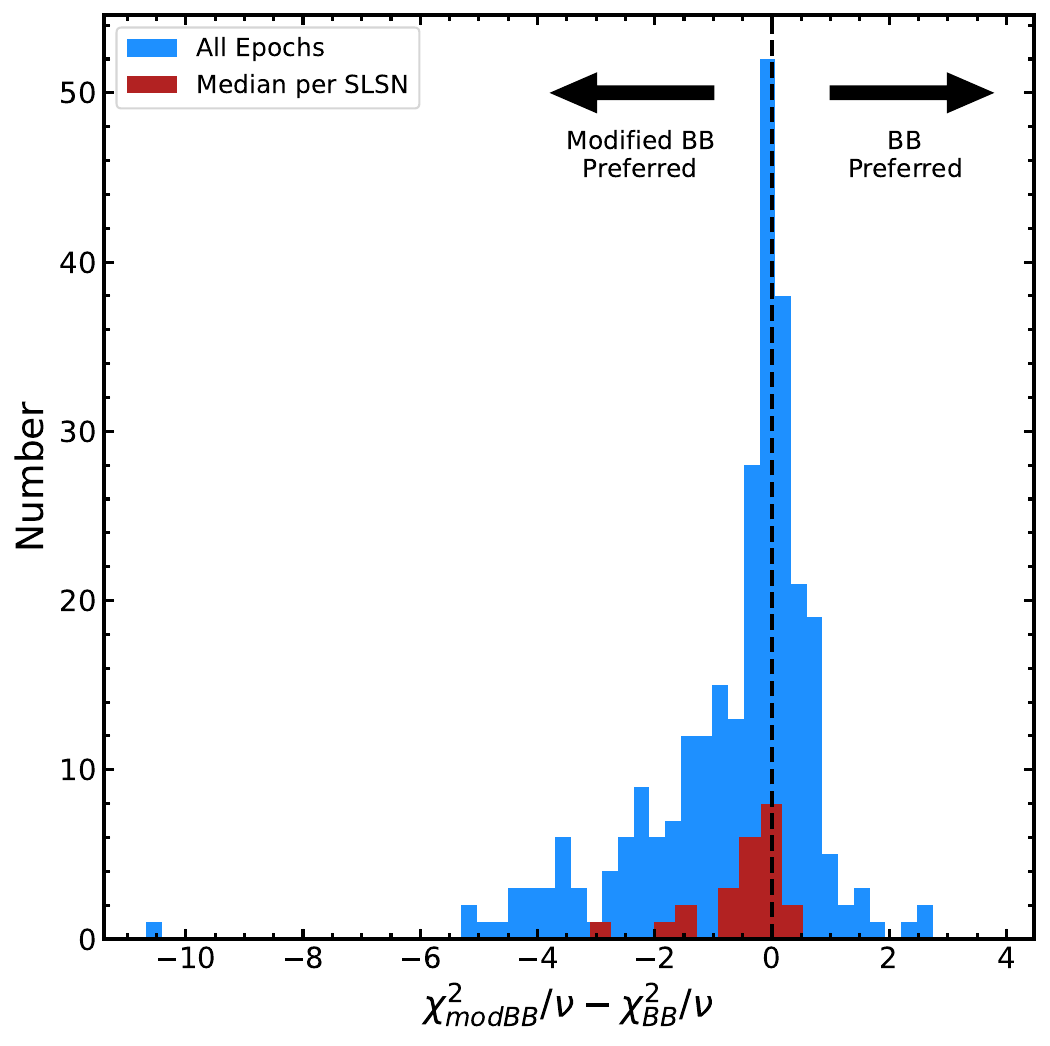}
 \caption{A comparison of the reduced $\chi^2$ values for modified blackbody and simple blackbody fits to the UV/optical SED of each SLSN. Negative differences indicate that a modified blackbody provides a better fit, whereas a positive difference indicates that a simple blackbody is preferred. The reduced $\chi^2$ differences for each \swift UVOT epoch are shown in the blue histogram, while the red histogram shows the median difference per object.}
 \label{fig:modBB_BB_comp}
\end{figure}

In addition to the modified blackbody model used above, we also fit each SLSN with a simple blackbody to compare the two SED models for a sample of SLSNe well-observed in the UV. The results of this comparison are illustrated in Figure \ref{fig:modBB_BB_comp}. While many epochs are similarly well fit by either model, we find that a modified blackbody is generally preferred compared to a simple blackbody. When considering each \swift UVOT epoch, a modified blackbody model produces a lower $\chi^2/\nu$ than a simple blackbody in 62\% of cases. Similarly, when evaluating the median $\chi^2/\nu$ difference per object, 18 out of 23 SLSNe, or 78\% of our objects, prefer a modified blackbody fit as compared to a simple blackbody. This, combined with the direct measurements of SLSNe-I SEDs from UV spectroscopy, suggests that a simple blackbody does not provide a sufficient description of the UV emission from SLSNe-I.

\section{Magnetar Modeling with \textsc{MOSFiT}} \label{sec:mosfit}

Beyond the modified blackbody fits to the SEDs of the SLSNe-I in our sample, we want to estimate physical parameters of the supernova explosion. One of the most commonly used models for SLSNe-I is the magnetar model \citep[e.g.,][]{kasen10, woosley10, nicholl17c}. In such a model, a rapidly spinning neutron star with a large magnetic field (i.e., a magnetar) injects energy from its spin-down into the supernova ejecta. When the timescales of the magnetar spin-down and diffusion time within the ejecta are well-matched, this powers transient emission significantly brighter than a typical core-collapse supernova \citep{kasen10, woosley10}.

Here, we use the Modular Open Source Fitter for Transients \citep[MOSFiT; ][]{guillochon18} to fit the observed emission from our SLSNe. In particular, we use the \textsc{slsn} model developed by \citet{nicholl17c}. The MOSFiT \textsc{slsn} model operates by initializing a magnetar central engine which spins down with time \citep[e.g.,][]{chatzopoulos12, inserra13}. This spin-down energy is then diffused through the ejecta \citep[e.g.,][]{arnett82, inserra13, wang15} to calculate a bolometric luminosity. MOSFiT then generates a temperature and radius from the physical parameters to create a transient SED. For the \textsc{slsn} model, the SED is a modified blackbody with a linear UV flux suppression at wavelengths shorter than 3000 \AA. Unlike many other models, MOSFiT compares directly to the multi-band photometry rather than pre-computing a bolometric light curve and then fitting a magnetar model.

In addition to the \swift UVOT data presented for our sample of SLSNe, we also obtained optical photometry of these events to better sample their light curve evolution. The optical data consisted of survey photometry from the Asteroid Terrestrial-impact Last Alert System \citep[ATLAS;][]{tonry18, smith20}, Gaia \citep{wyrzykowski12}, the Panoramic Survey Telescope and Rapid Response System \citep[Pan-STARRS;][]{chambers16}, the Zwicky Transient Facility \citep[ZTF;][]{bellm19}, and/or ground-based follow-up photometry from the literature as appropriate. The long-baseline optical data, combined with the UV data near peak, is crucial as the decline rate is important for determining the magnetic field strength \citep{nicholl17c}. 

We ran MOSFiT for each of our SLSNe using MCMC sampling with 300 walkers and the default priors for the \textsc{slsn} model. We recorded the bolometric luminosity for each chain as a function of time using the \textsc{dense\_luminosities} and \textsc{dense\_times} flags with additional temporal coverage of 1000 days on either end of the observed data. To ensure self-consistency between our modified blackbody fits and the MOSFiT fits, we specified the redshift and luminosity distance for each SN. Nominally, we ran each MOSFiT model until convergence at a Potential Scale Reduction Factor (PSRF) value of 1.2 \citep{gelman92, brooks98, nicholl17c}. In some cases, the runtime on the fits was prohibitively long, so we terminated them after reaching a PSRF value below 1.3. The mean and median PSRF for the full sample were 1.25. We excluded OGLE16dmu from our MOSFiT modeling as it has only marginal coverage in the \swift $U$ and $UVW1$ bands and no published optical light curves. The median and 1$\sigma$ uncertainties on key magnetar parameters from our MOSFiT modeling are given in Table \ref{tab:MOSFIT_table}.

\begin{deluxetable*}{cccccccc}[h]
\tablecaption{MOSFiT Medians and 1$\sigma$ Uncertainties}
\tablehead{
\colhead{Object} &
\colhead{TNS ID} &
\colhead{B$_{\perp}$} &
\colhead{P} &
\colhead{M$_{NS}$} &
\colhead{M$_{ej}$} &
\colhead{v$_{ej}$} & \\
\colhead{} &
\colhead{} &
\colhead{(10$^{14}$ \textrm{ G})} &
\colhead{(ms)} &
\colhead{(M$_\odot$)} &
\colhead{(M$_\odot$)} &
\colhead{(10$^{3}$ \textrm{ km s}$^{-1}$)} &
}
\startdata
ATLAS18unu &  SN2018ibb & $0.31^{+0.85}_{-0.20}$ & $2.52^{+0.29}_{-0.50}$ & $1.66^{+0.24}_{-0.39}$ & $18.62^{+3.77}_{-2.40}$ & $5.05^{+0.07}_{-0.04}$ \\
ATLAS18yff &  SN2018hti & $0.53^{+1.33}_{-0.29}$ & $1.94^{+0.29}_{-0.42}$ & $1.64^{+0.27}_{-0.38}$ & $7.41^{+1.50}_{-0.96}$ & $7.13^{+0.16}_{-0.20}$ \\
ATLAS19ine &  SN2019enz & $3.03^{+2.01}_{-1.20}$ & $1.65^{+0.73}_{-0.46}$ & $1.65^{+0.23}_{-0.33}$ & $4.37^{+2.24}_{-1.35}$ & $11.37^{+0.62}_{-0.69}$ \\
ATLAS19prf &  SN2019lsq & $0.80^{+1.65}_{-0.38}$ & $4.92^{+0.63}_{-0.77}$ & $1.60^{+0.30}_{-0.32}$ & $2.57^{+1.70}_{-0.95}$ & $6.70^{+0.21}_{-0.22}$ \\
ATLAS19ynd &  SN2019szu & $0.65^{+1.70}_{-0.35}$ & $1.34^{+0.46}_{-0.23}$ & $1.74^{+0.18}_{-0.32}$ & $46.77^{+14.89}_{-11.29}$ & $5.10^{+0.17}_{-0.07}$ \\
ATLAS20xqi &  SN2020rmv & $0.63^{+1.40}_{-0.39}$ & $1.68^{+1.08}_{-0.51}$ & $1.67^{+0.23}_{-0.33}$ & $23.99^{+11.49}_{-13.76}$ & $6.13^{+0.41}_{-0.39}$ \\
ATLAS20zst &  SN2020tcw & $1.21^{+1.25}_{-0.42}$ & $1.20^{+0.20}_{-0.13}$ & $1.73^{+0.19}_{-0.28}$ & $7.08^{+1.24}_{-0.91}$ & $16.29^{+1.07}_{-1.12}$ \\
Gaia16apd  &  SN2016eay & $1.20^{+1.61}_{-0.62}$ & $2.31^{+0.41}_{-0.42}$ & $1.64^{+0.26}_{-0.36}$ & $5.89^{+1.70}_{-1.10}$ & $9.70^{+0.62}_{-0.63}$ \\
Gaia17biu  &  SN2017egm & $1.42^{+1.84}_{-0.72}$ & $4.32^{+0.63}_{-0.92}$ & $1.63^{+0.26}_{-0.37}$ & $3.16^{+1.85}_{-0.59}$ & $6.28^{+0.32}_{-0.35}$ \\
Gaia17cbp  &  SN2017gci & $1.64^{+1.99}_{-0.70}$ & $1.20^{+0.28}_{-0.16}$ & $1.74^{+0.19}_{-0.29}$ & $9.55^{+5.24}_{-2.31}$ & $10.88^{+1.55}_{-1.14}$ \\
Gaia18beg  &  SN2018bgv & $2.21^{+2.21}_{-0.84}$ & $8.16^{+1.19}_{-1.42}$ & $1.54^{+0.29}_{-0.25}$ & $1.26^{+0.65}_{-0.33}$ & $8.59^{+1.63}_{-1.58}$ \\
LSQ14mo    &  --- & $5.42^{+1.50}_{-1.75}$ & $2.06^{+0.95}_{-0.55}$ & $1.58^{+0.29}_{-0.35}$ & $3.09^{+0.80}_{-0.64}$ & $15.77^{+1.43}_{-1.20}$ \\
PS15ae     &  SN2015bn & $0.44^{+0.90}_{-0.22}$ & $1.49^{+0.23}_{-0.25}$ & $1.63^{+0.30}_{-0.37}$ & $12.59^{+3.63}_{-1.62}$ & $8.38^{+0.56}_{-0.60}$ \\
PS16aqv    &  SN2016ard & $0.65^{+1.40}_{-0.37}$ & $3.08^{+0.54}_{-0.55}$ & $1.66^{+0.24}_{-0.34}$ & $2.14^{+0.81}_{-0.66}$ & $10.47^{+1.14}_{-0.77}$ \\
PS16dnq    &  SN2016els & $2.24^{+2.27}_{-0.87}$ & $1.35^{+0.51}_{-0.26}$ & $1.71^{+0.20}_{-0.35}$ & $7.59^{+3.38}_{-1.96}$ & $9.59^{+2.15}_{-1.64}$ \\
PS22bca    &  SN2021ahpl & $0.62^{+1.61}_{-0.37}$ & $8.23^{+1.12}_{-1.52}$ & $1.62^{+0.24}_{-0.35}$ & $4.79^{+2.46}_{-1.62}$ & $5.41^{+0.61}_{-0.31}$ \\
ZTF19aawfbtg  &  SN2019hge & $1.86^{+1.73}_{-0.65}$ & $4.91^{+1.17}_{-1.01}$ & $1.64^{+0.24}_{-0.37}$ & $8.71^{+2.77}_{-1.63}$ & $5.05^{+0.09}_{-0.04}$ \\
ZTF19abpbopt  &  SN2019neq  & $1.42^{+1.88}_{-0.69}$ & $4.91^{+0.54}_{-0.86}$ & $1.66^{+0.25}_{-0.37}$ & $2.19^{+0.38}_{-0.19}$ & $11.72^{+0.29}_{-0.29}$ \\
ZTF20abobpcb  &  SN2020qlb & $0.42^{+0.62}_{-0.28}$ & $1.24^{+0.16}_{-0.14}$ & $1.76^{+0.17}_{-0.26}$ & $27.54^{+4.08}_{-3.55}$ & $8.31^{+0.18}_{-0.20}$ \\
ZTF20acphdcg  &  SN2020znr & $0.46^{+1.33}_{-0.27}$ & $2.79^{+0.25}_{-0.50}$ & $1.70^{+0.20}_{-0.38}$ & $16.22^{+1.98}_{-1.76}$ & $5.18^{+0.10}_{-0.10}$ \\
ZTF20acpyldh  &  SN2020abjc & $0.10^{+0.34}_{-0.08}$ & $2.37^{+0.40}_{-0.51}$ & $1.67^{+0.23}_{-0.34}$ & $20.89^{+4.23}_{-3.91}$ & $5.31^{+0.53}_{-0.22}$ \\
ZTF21aaarmti  &  SN2021ek & $1.92^{+2.21}_{-0.94}$ & $3.79^{+1.18}_{-1.69}$ & $1.60^{+0.32}_{-0.41}$ & $4.57^{+2.51}_{-1.62}$ & $8.14^{+1.16}_{-1.00}$ \\
ZTF21abaiono  &  SN2021lwz & $5.52^{+1.79}_{-1.70}$ & $8.35^{+1.27}_{-1.33}$ & $1.40^{+0.29}_{-0.24}$ & $0.89^{+0.55}_{-0.30}$ & $8.42^{+0.72}_{-0.52}$ \\
\enddata 
\tablecomments{Median values and 1$\sigma$ uncertainties for key MOSFiT parameters.}
\label{tab:MOSFIT_table}
\end{deluxetable*}

\subsection{Peak UV/optical Luminosity and Emitted Energy}

\begin{figure*}
\centering
 \includegraphics[height=0.48\textwidth]{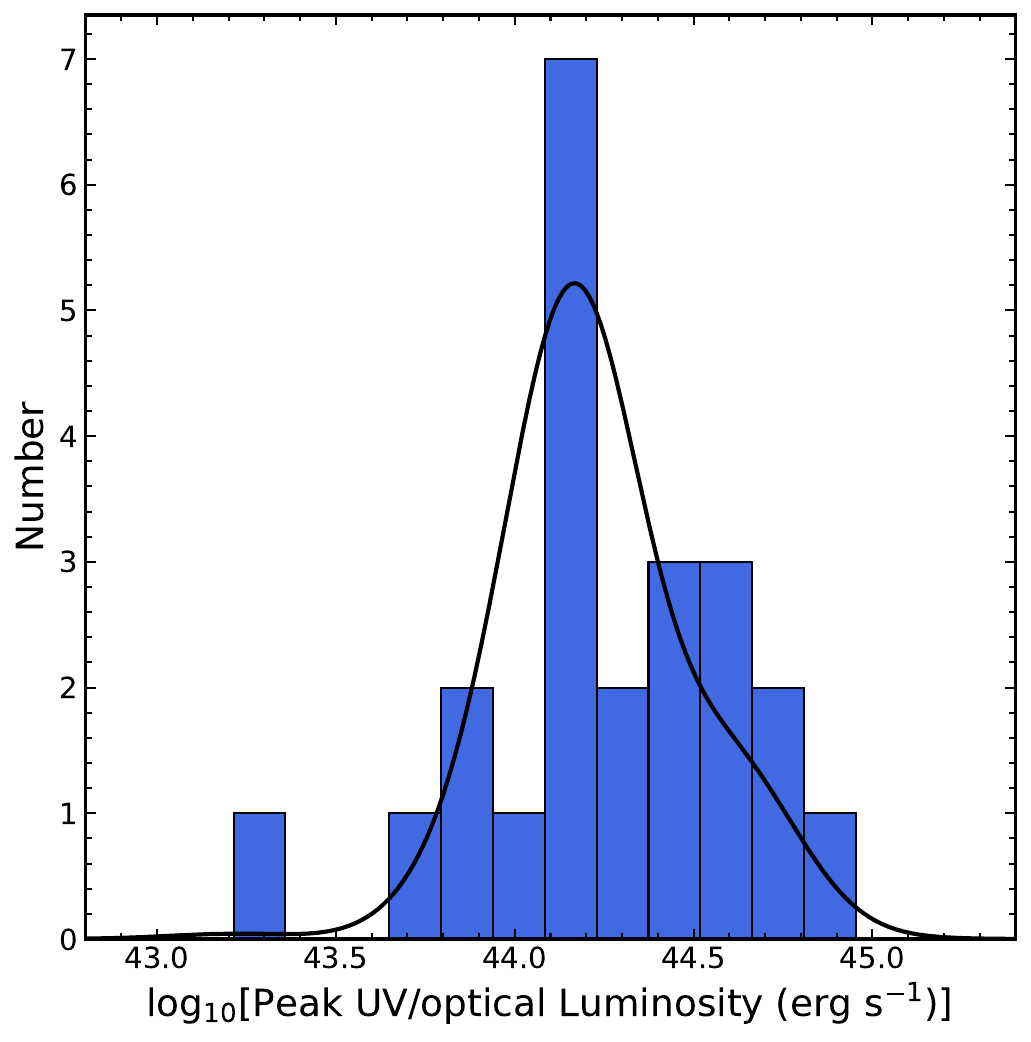}
 \includegraphics[height=0.48\textwidth]{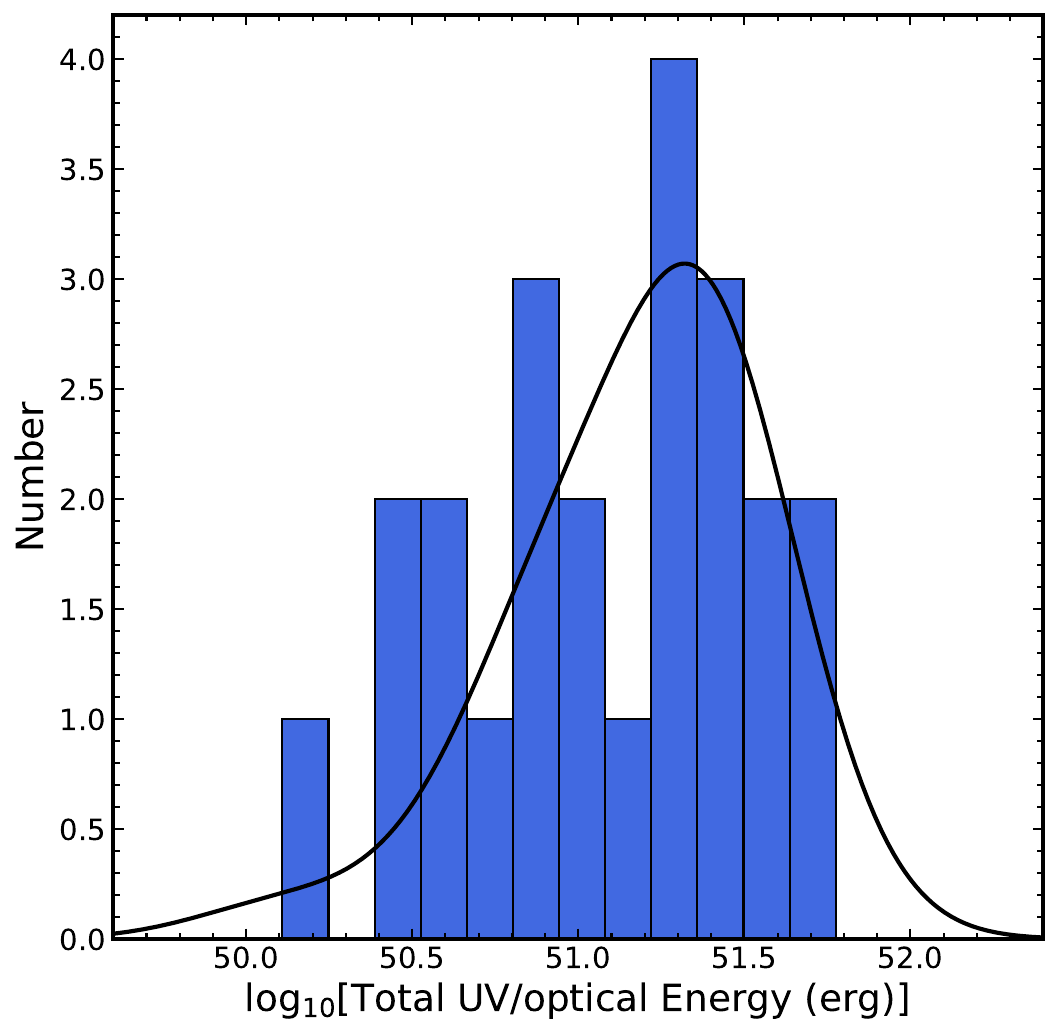}
 \caption{Histograms of peak UV/optical luminosity and radiative energy computed from the MOSFiT outputs. Shown in black are KDEs of the radius and temperature distributions normalized to the sample size. The individual SN contributions to the KDEs are weighted by the inverse square of the estimated uncertainty on the peak luminosity and energy.}
 \label{fig:LE_hist}
\end{figure*}

In Figure \ref{fig:LE_hist} we show the distribution of peak luminosities and emitted energies for our sample of SLSNe along with corresponding KDEs. Given the dense temporal sampling, we computed the peak luminosity by taking the maximum value of the bolometric luminosity curves from MOSFiT. We took the median value as the peak luminosity and the 16th and 84th percentiles as the 1$\sigma$ bounds on the peak luminosity. The distributions of peak luminosities from MOSFiT, as compared to the peak luminosities from our modified blackbody fits, are similar. The luminosity distribution peaks at $1.5 \times 10^{44}$ erg s$^{-1}$, with a slight bump near $5 \times 10^{44}$ erg s$^{-1}$. The distribution spans $(2-90) \times 10^{43}$ erg s$^{-1}$, fully consistent with the luminosity distribution seen in previous studies \citep{lunnan18, decia18, angus19, chen23a}. The median peak luminosity is also similar to other studies, although this sample appears to have fewer low-luminosity SLSNe than some other samples \citep{angus19}. This is likely a result of the targeted nature of \swift follow-up and the tendency for brighter SLSNe to be bluer \citep{chen23a}.

To calculate the radiative energy emitted, we integrated the bolometric luminosity over time and again took the median as the radiative energy with the 16th and 84th percentiles as the 1$\sigma$ bounds. The radiative energy distribution ranges from $(1 - 60) \times 10^{50}$ erg, with a peak at $2 \times 10^{51}$ erg. This is largely consistent with the emitted energies of the sample of SLSNe from  \citep{lunnan18}, especially since that sample does not cover the full SN light curve and may not account for all of the UV emission.

\begin{figure*}
\centering
 \includegraphics[height=0.4\textwidth]{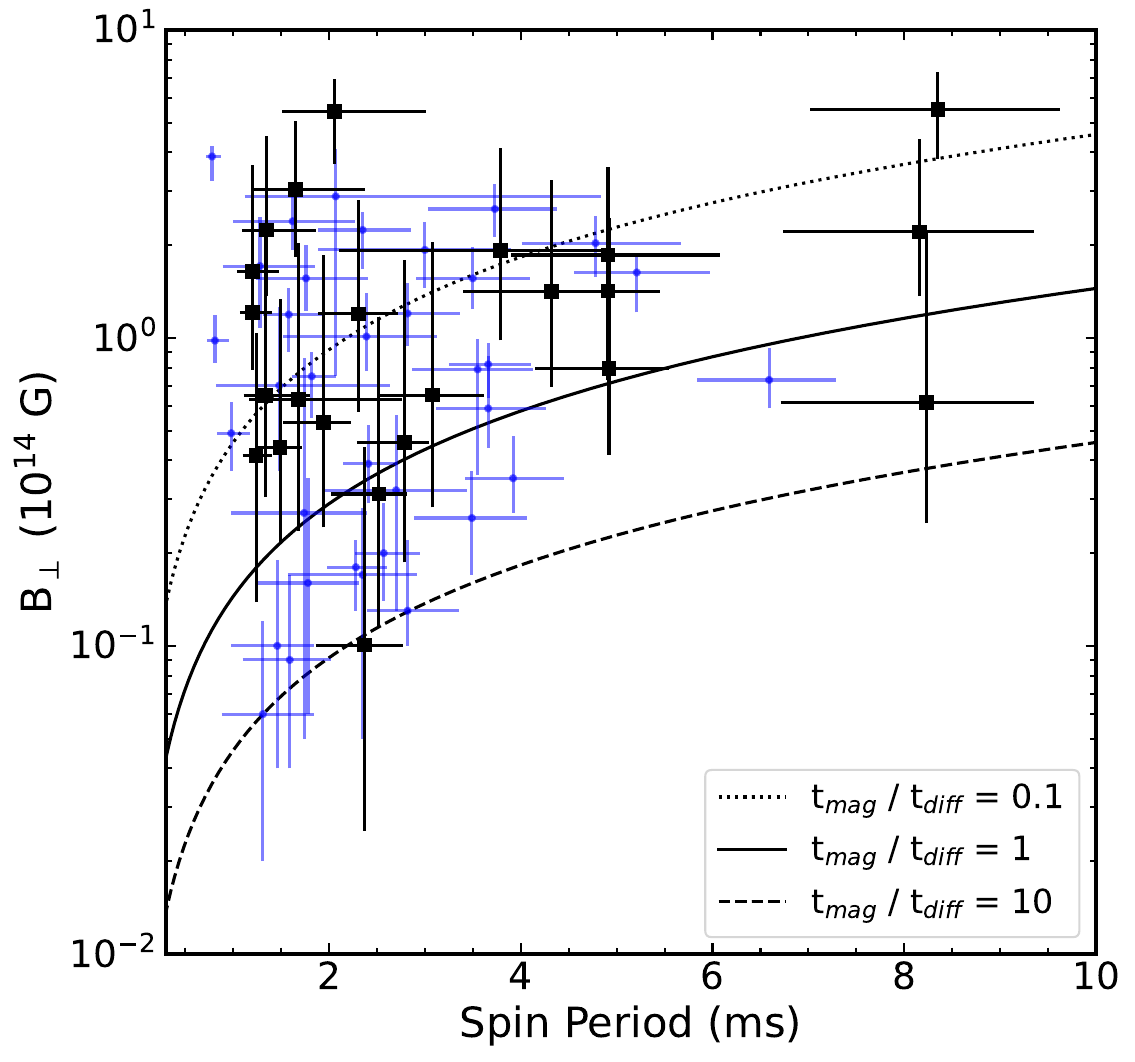}
 \includegraphics[height=0.4\textwidth]{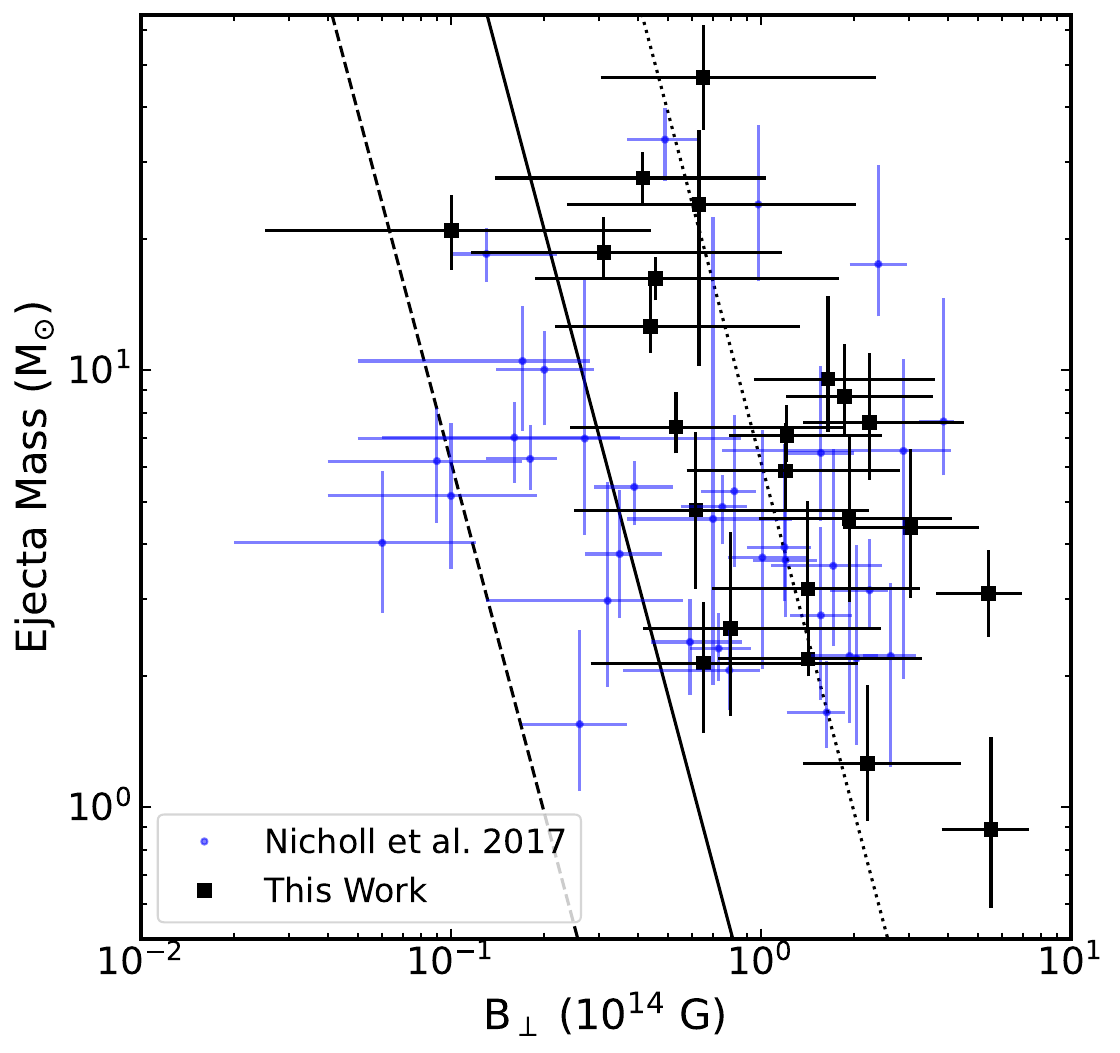}
 \includegraphics[height=0.4\textwidth]{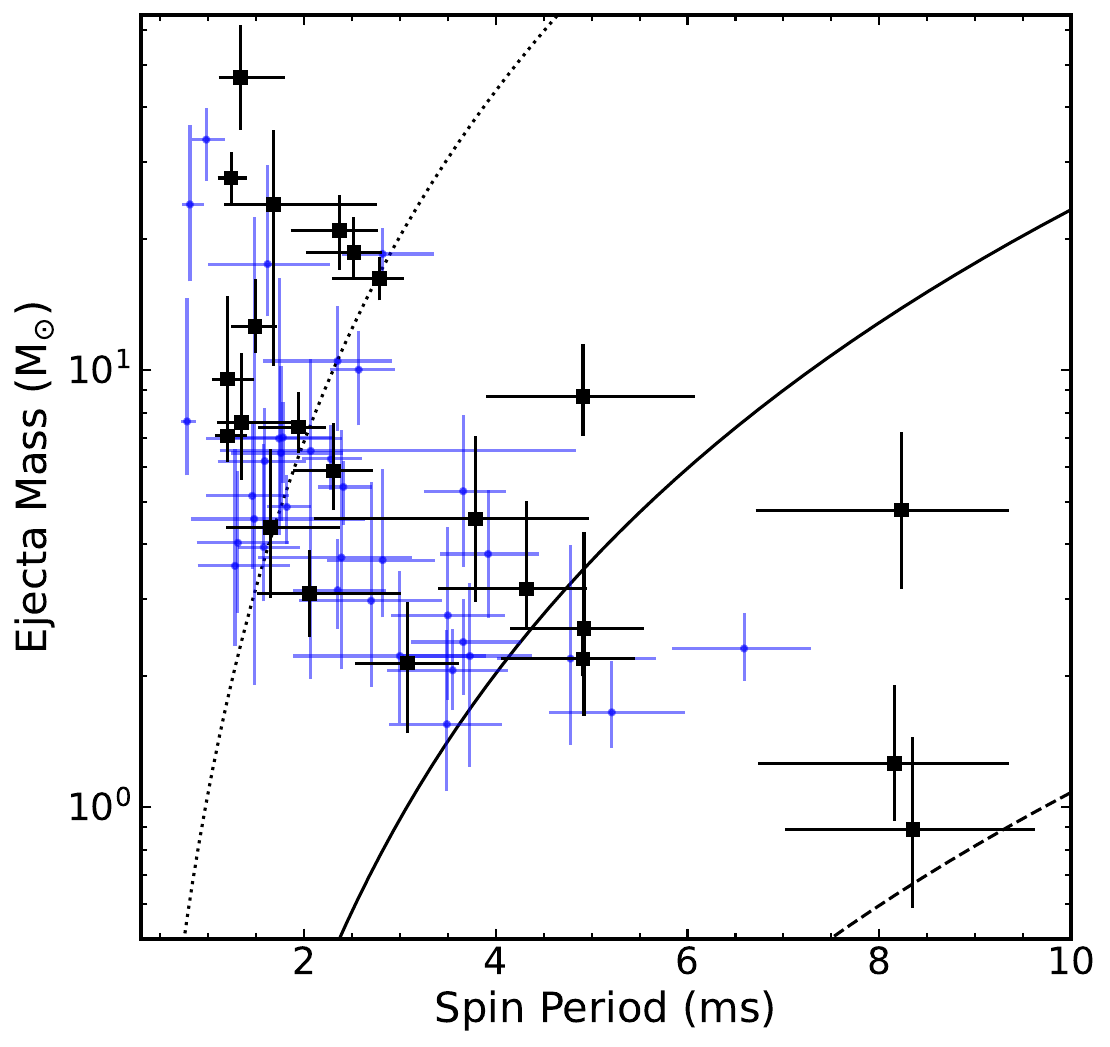}
 \includegraphics[height=0.4\textwidth]{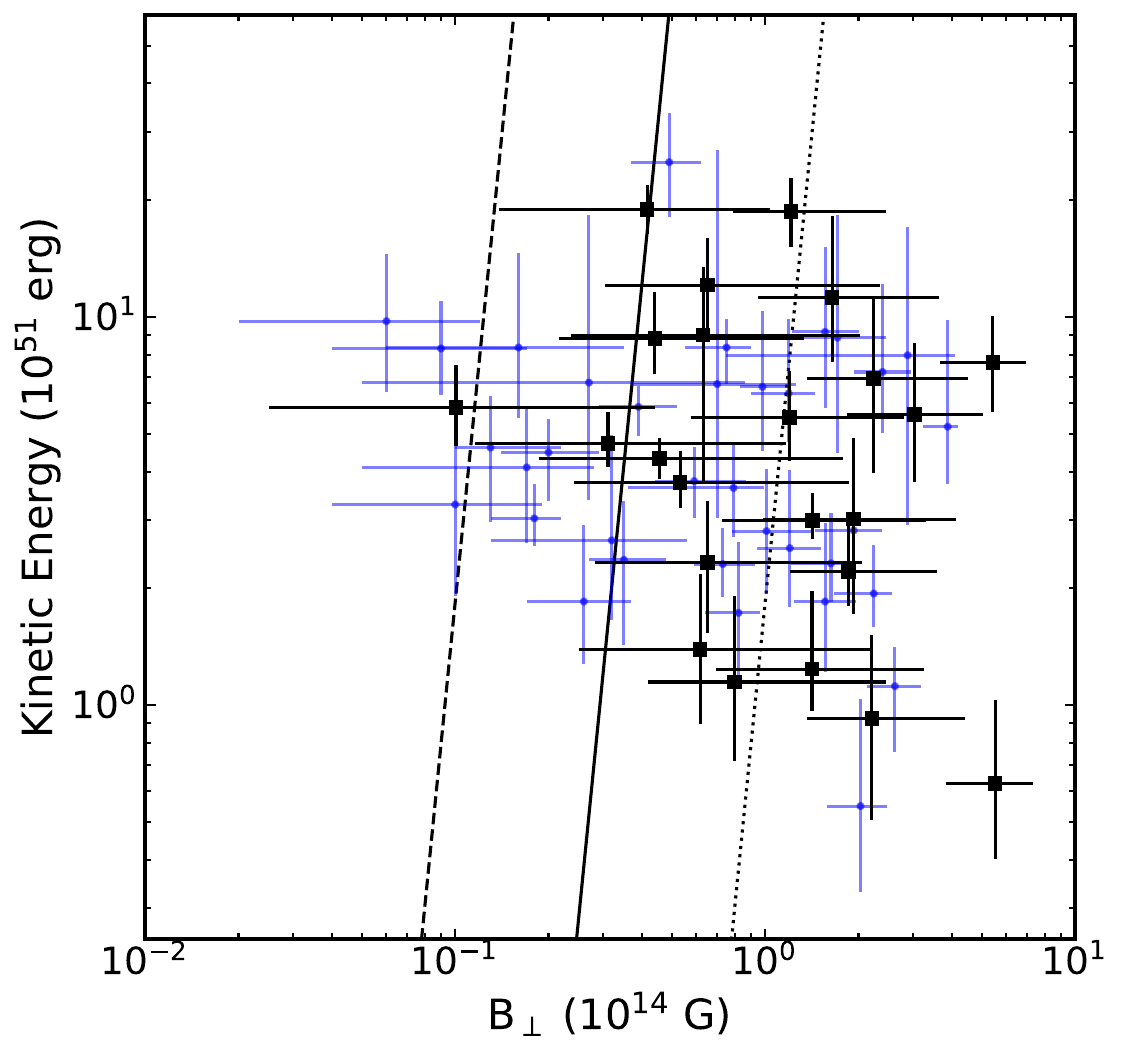}
 \includegraphics[height=0.4\textwidth]{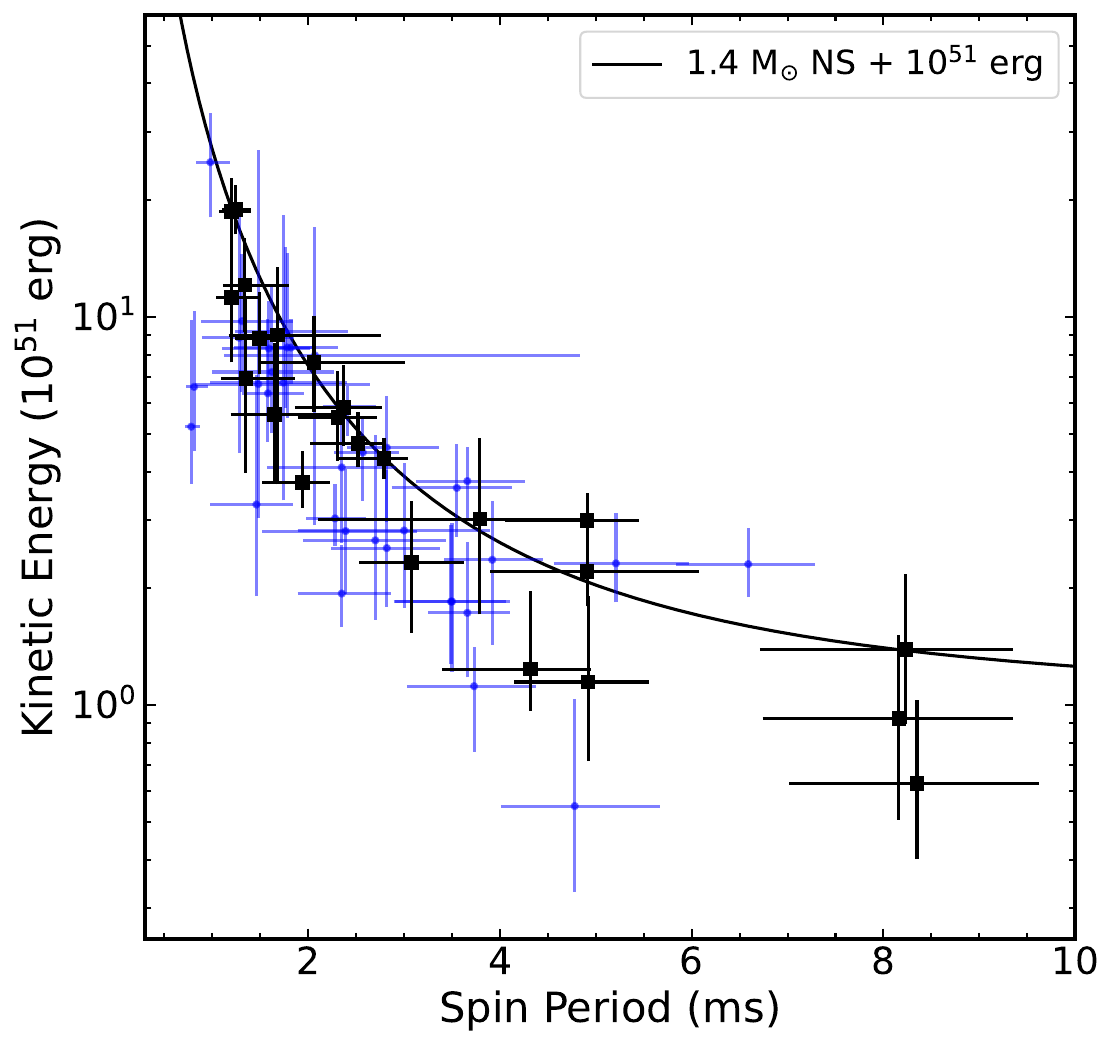}
 \includegraphics[height=0.4\textwidth]{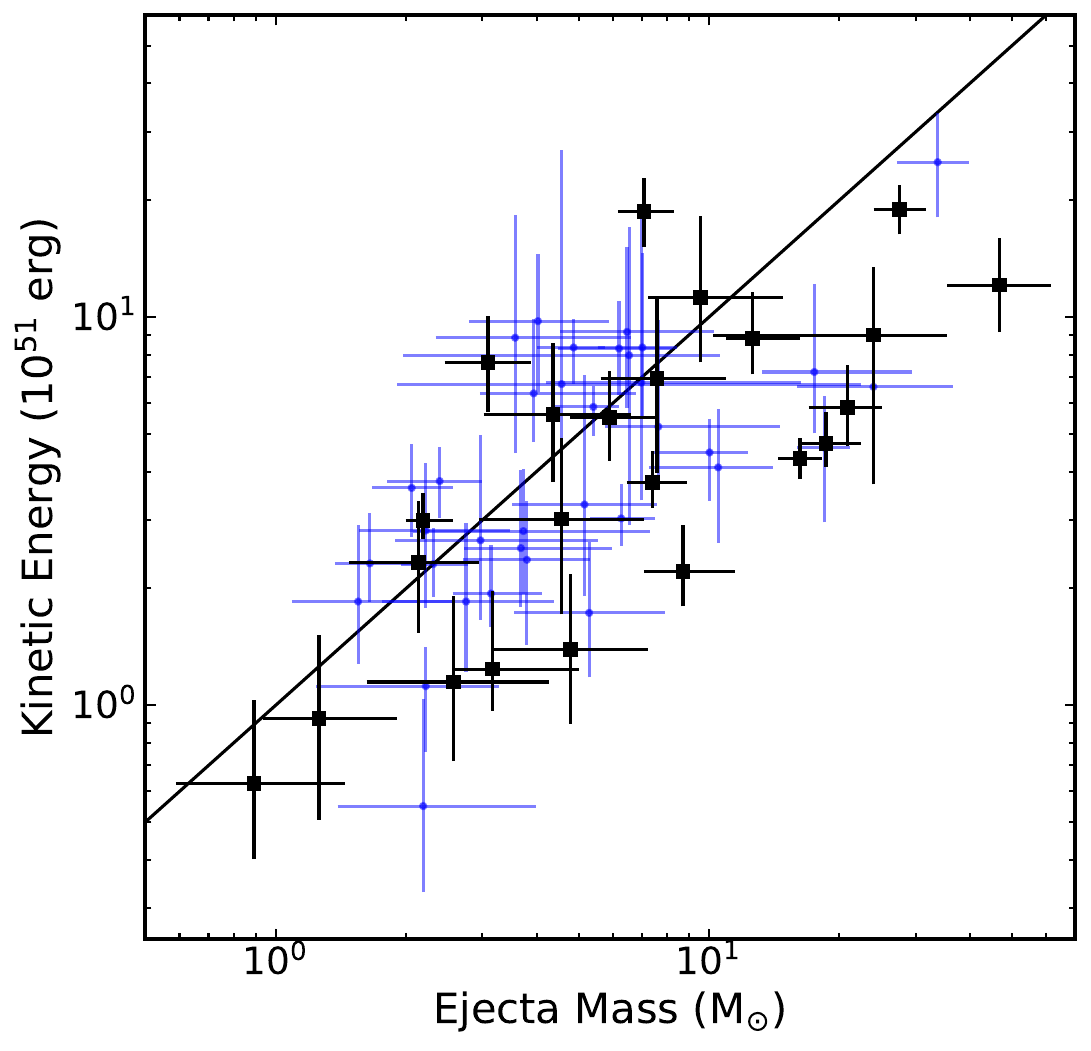}
 \caption{Key physical parameters (spin period, magnetic field, ejecta mass, and kinetic energy) for the SLSNe in this sample (black squares) and a comparison sample from \citet[][blue circles]{nicholl17c}. The lines in the first four panels are lines of constant ratio between the magnetar spin-down timescale and the diffusion timescale. The line in the bottom left panel is a sum of the rotational energy of the NS and a characteristic SN energy. The line in the bottom right panel is a 1:1 line.}
 \label{fig:MOSFiT}
\end{figure*}

\subsection{Estimated Physical Parameters}

Beyond estimates of luminosity and energy, we used our MOSFiT modeling to estimate key physical parameters of the newly-formed neutron star and the supernova ejecta. In Figure \ref{fig:MOSFiT}, we show various parameter combinations from MOSFiT along with a comparison set of SLSNe from \citet{nicholl17c}. We also plot several lines representing different ratios of the magnetar spin-down timescale (t$_{\textrm{mag}}$) and diffusion timescale (t$_{\textrm{diff}}$), assuming the median parameters of \citet{nicholl17c} for the parameters not shown in a given panel. The magnetar spin-down timescale is given by

\begin{equation}
    t_{\textrm{mag}} \simeq \frac{P}{2\dot{P}}
\end{equation}

\noindent where P is the magnetar spin period and $\dot{P}$ is the period derivative \citep{ostriker71, nicholl17c}. The diffusion timescale is

\begin{equation}
    t_{\textrm{diff}} = \left( \frac{2 \kappa M_{\textrm{ej}}}{\beta c v_{\textrm{ej}}} \right)^{1/2}
\end{equation}

where $\kappa$ is the optical opacity, $M_{\textrm{ej}}$ is the ejecta mass, $\beta$ is a constant accounting for the density distribution of the ejecta, $c$ is the speed of light, and $v_{\textrm{ej}}$ is the ejecta velocity \citep{arnett82, nicholl17c}.

Overall, we find good agreement between our sample and the \citet{nicholl17c} sample for each of the parameters. Using a K-S test \citep{massey51}, the distributions for each of the key physical parameters are consistent between our sample and that of \citet{nicholl17c}. Our median and $1\sigma$ dispersion on key parameters are as follows: P $= 2.4^{+2.5}_{-1.0}$ ms, B$_{\perp} = 1.2^{+1.0}_{-0.7} \times 10^{14}$ G, M$_{ej} = 7.1^{+12.6}_{-4.7} M_{\odot}$, E$_{kin} = 4.7^{+5.3}_{-3.4} \times 10^{51}$ erg, fully consistent with previous studies \citep{nicholl17c, hsu21, chen23b}.

The top left panel compares the perpendicular magnetic field to the NS spin period. We find good agreement with previous samples, although slightly skewed towards higher magnetic field strengths. Regardless, the K-S test discussed above indicates no statistically significant difference in the distributions of magnetic field strength. In terms of the t$_{\textrm{mag}}$ / t$_{\textrm{diff}}$ ratio, the SNe tend to prefer a value below 1. The top right panel compares the ejecta mass with the magnetic field. Interestingly, the sources in our sample lie neatly on the t$_{\textrm{mag}}$ / t$_{\textrm{diff}} \sim$0.1 line. However, several sources in the \citet{nicholl17c} sample lie off this line. There also appears to be a moderately significant anti-correlation between the ejecta mass and magnetic field, with Kendall $\tau = -0.47$ and corresponding p-value of $1.3 \times 10^{-3}$. However, this may simply be the result of observational bias, as SLSNe with lower ejecta masses and weaker magnetic fields tend to be less luminous.

The middle left panel compares the ejecta mass and spin period. Our sample is similar to that of \citet{nicholl17c}, but with higher scatter. Regardless, we confirm the anti-correlation noted by previous studies \citep{nicholl17c, blanchard20, hsu21, chen23b}. The middle right panel compares the kinetic energy with the magnetic field strength. Here we have computed the kinetic energy as $E_k = 1/2 M_{ej} v_{phot}^{2}$ \citep{nicholl17c, margalit18}. We note that under the assumption of a homologous density profile, this relationship is instead $E_k = 3/10 M_{ej} v_{phot}^{2}$, although such a difference is unimportant for this study. In neither the \citet{nicholl17c} sample nor our sample do we find any SLSN that favors t$_{\textrm{mag}}$ / t$_{\textrm{diff}} > 10$ in all parameter comparisons.

The bottom left panel compares the kinetic energy with the NS spin period. The plotted line is the sum of the NS spin energy and a characteristic $10^{51}$ energy for supernovae. Again, we find good agreement with previous work. In the bottom right panel, we compare kinetic energy and ejecta mass along with a one-to-one line. In both the \citet{nicholl17c} sample and our sample, the ejecta mass and kinetic energy scale together as expected.

\begin{figure}
\centering
 \includegraphics[width=0.49\textwidth]{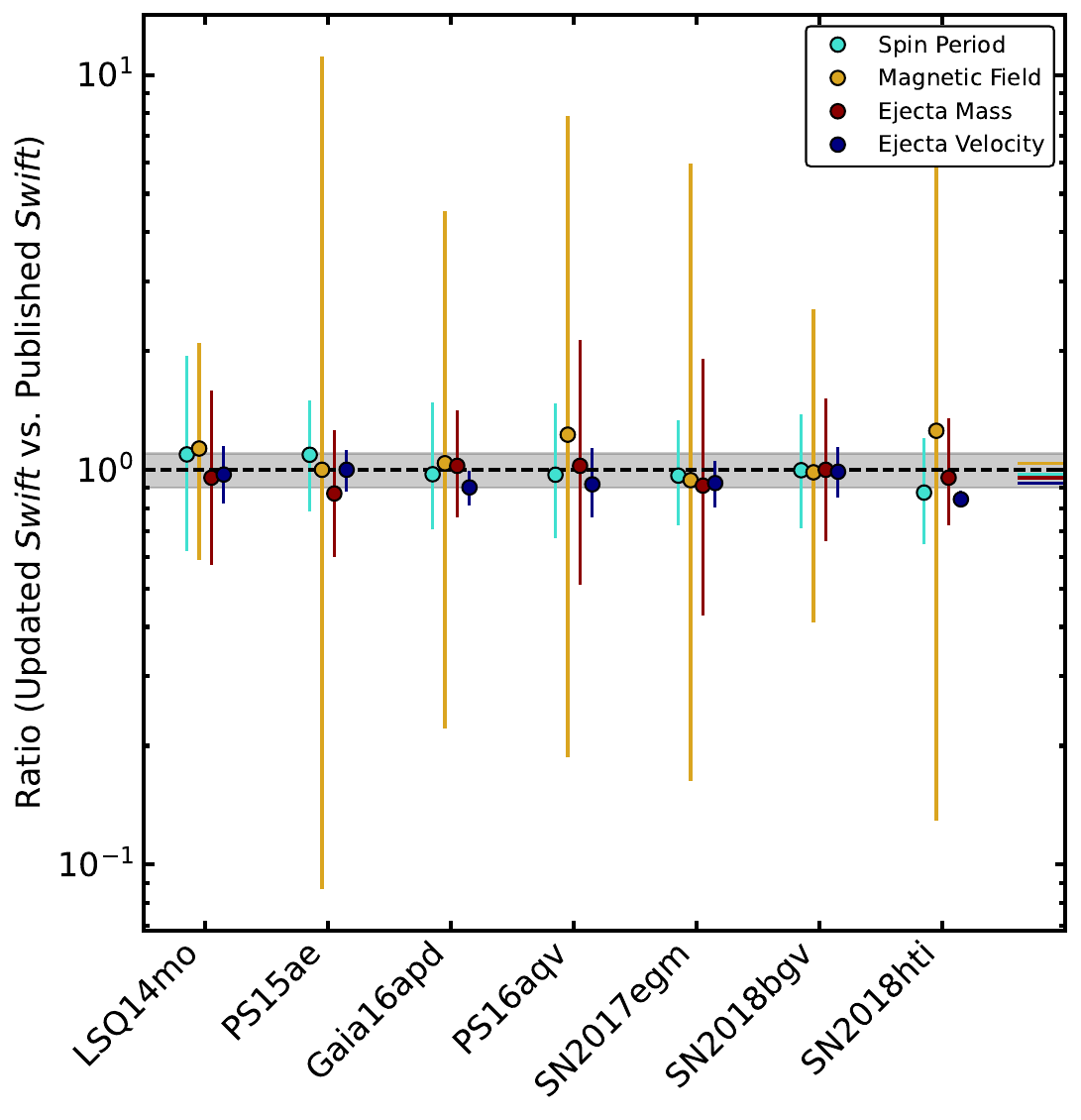}
 \caption{Ratio of physical parameters (spin period in teal, magnetic field in gold, ejecta mass in red, and ejecta velocity in navy) estimated from MOSFiT fits of the updated \swift photometry as compared to MOSFiT fits of published, pre-correction, \swift photometry. The black dashed line is a ratio of one, with the gray shading indicating 10\%. The colored lines on the right side are the median ratio for the corresponding physical parameters, all consistent within 10\%.}
 \label{fig:MOSFiT_comp}
\end{figure}

\subsection{Effect of Updated \swift Reductions}

The \swift UVOT photometry provides strong constraints on the SLSN temperature. Therefore, we ask what effect the updated \swift data has on our inferred MOSFiT parameters. To test this, we fit several SNe that have published pre-correction \swift and compared them to our fits including the updated \swift data. These results are shown in Figure \ref{fig:MOSFiT_comp}. Across all of the key parameters, we find good agreement between the values from fits including published and updated \textit{Swift}, with all having median ratios less than 10\%. The ejecta mass and ejecta velocity are the most different, expected as the different temperature constraints affect the diffusion timescale \citep{nicholl17c}. 

The lack of stark difference in inferred parameters may not be particularly surprising given the high redshifts of many SLSNe. Each of our SLSNe has observer-frame UV data, whereas this is not true for a large majority of the \citet{nicholl17c} sample. However, when accounting for the redshift, 60\% of the \citet{nicholl17c} sample has a rest-frame wavelength of $< 3000$ \AA\ for the bluest bandpass, and all have a bluest filter with a rest-frame wavelength blue-ward of \swift $B$. Therefore, given the median peak temperature of $\sim$11,000 K, the temperature may still be reasonably well-constrained even without observer-frame UV data.

\section{Correlations Between SLSN Parameters and Radiative Emission} \label{sec:energy_corr}

\begin{figure*}
\centering
 \includegraphics[width=0.48\textwidth]{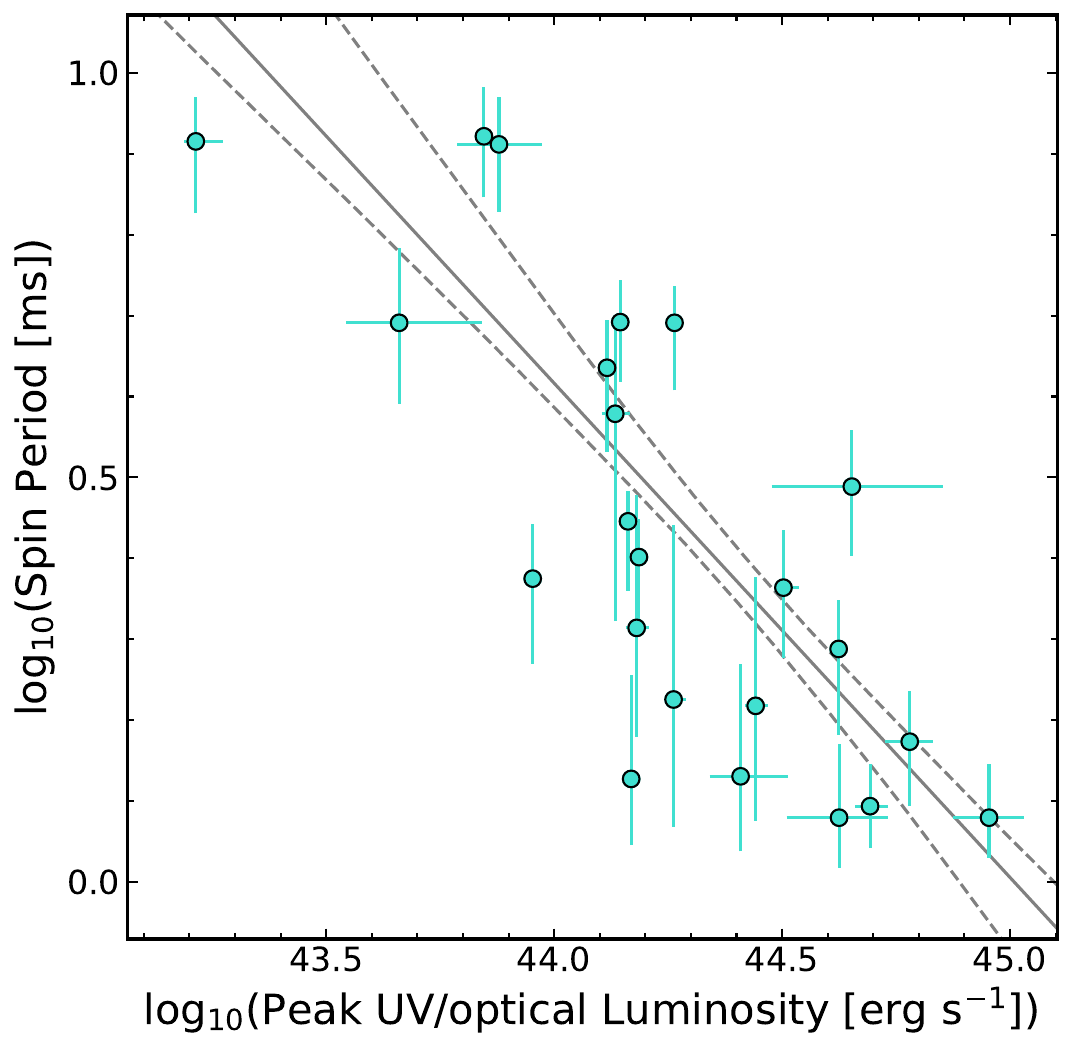}
 \includegraphics[width=0.48\textwidth]{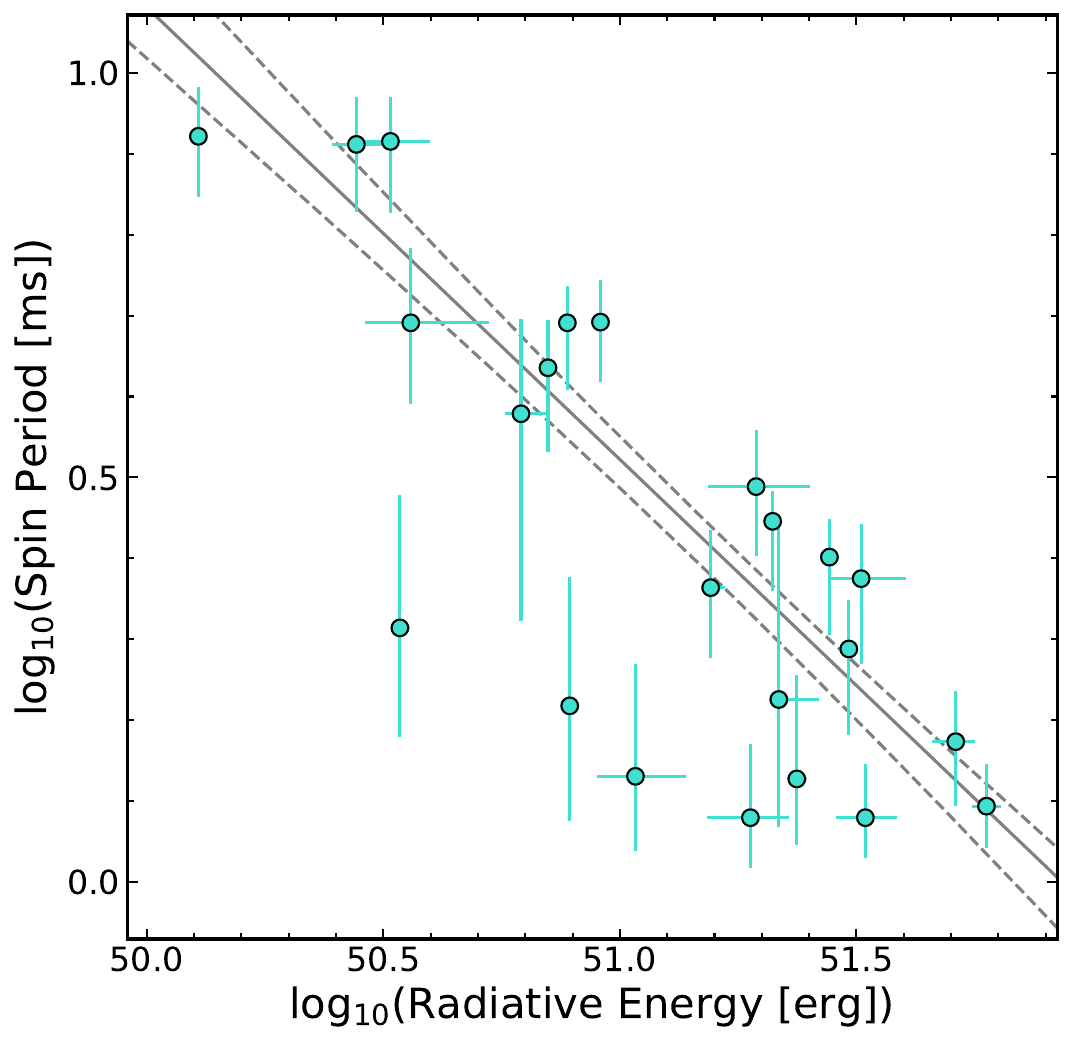}
 \includegraphics[width=0.48\textwidth]{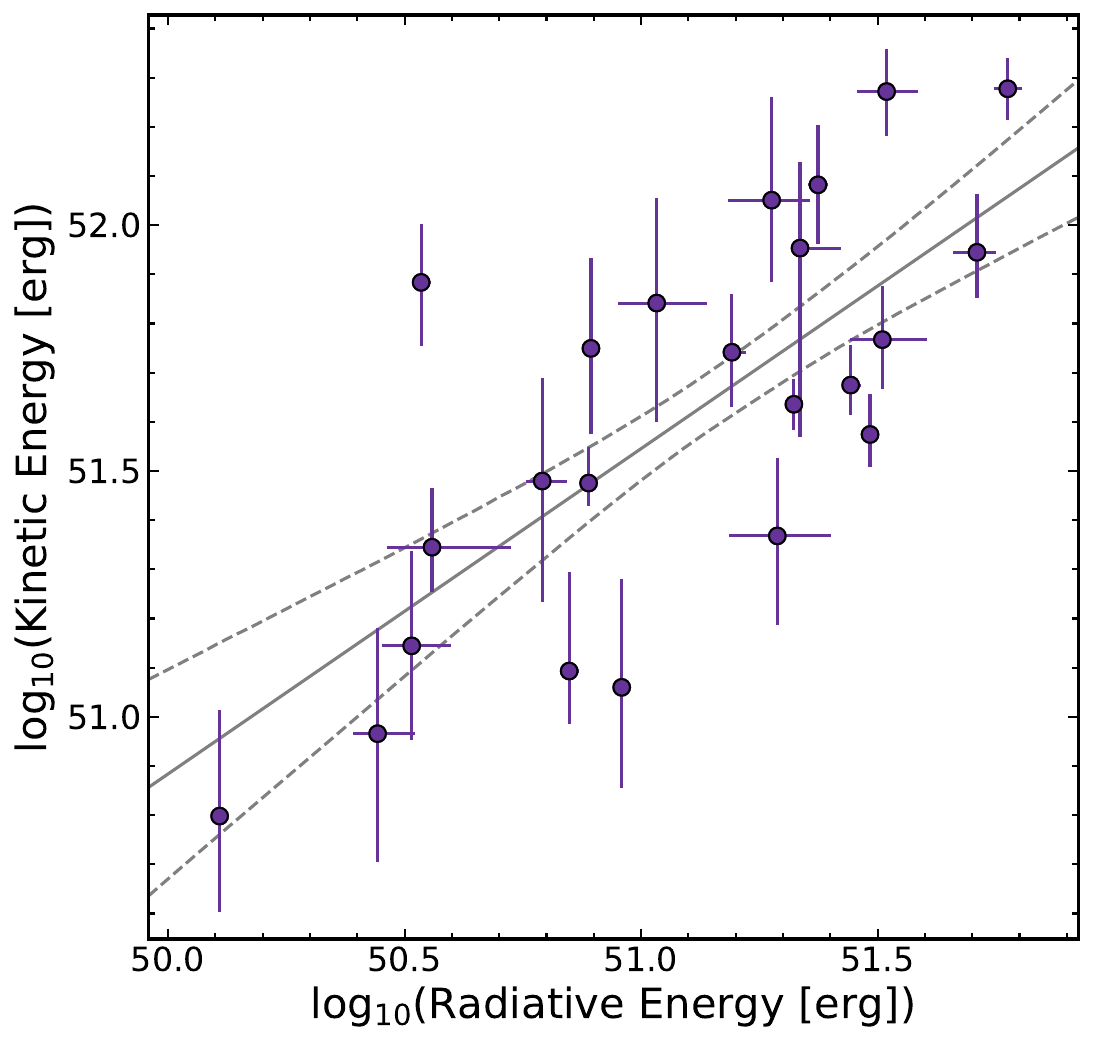}
 \includegraphics[width=0.48\textwidth]{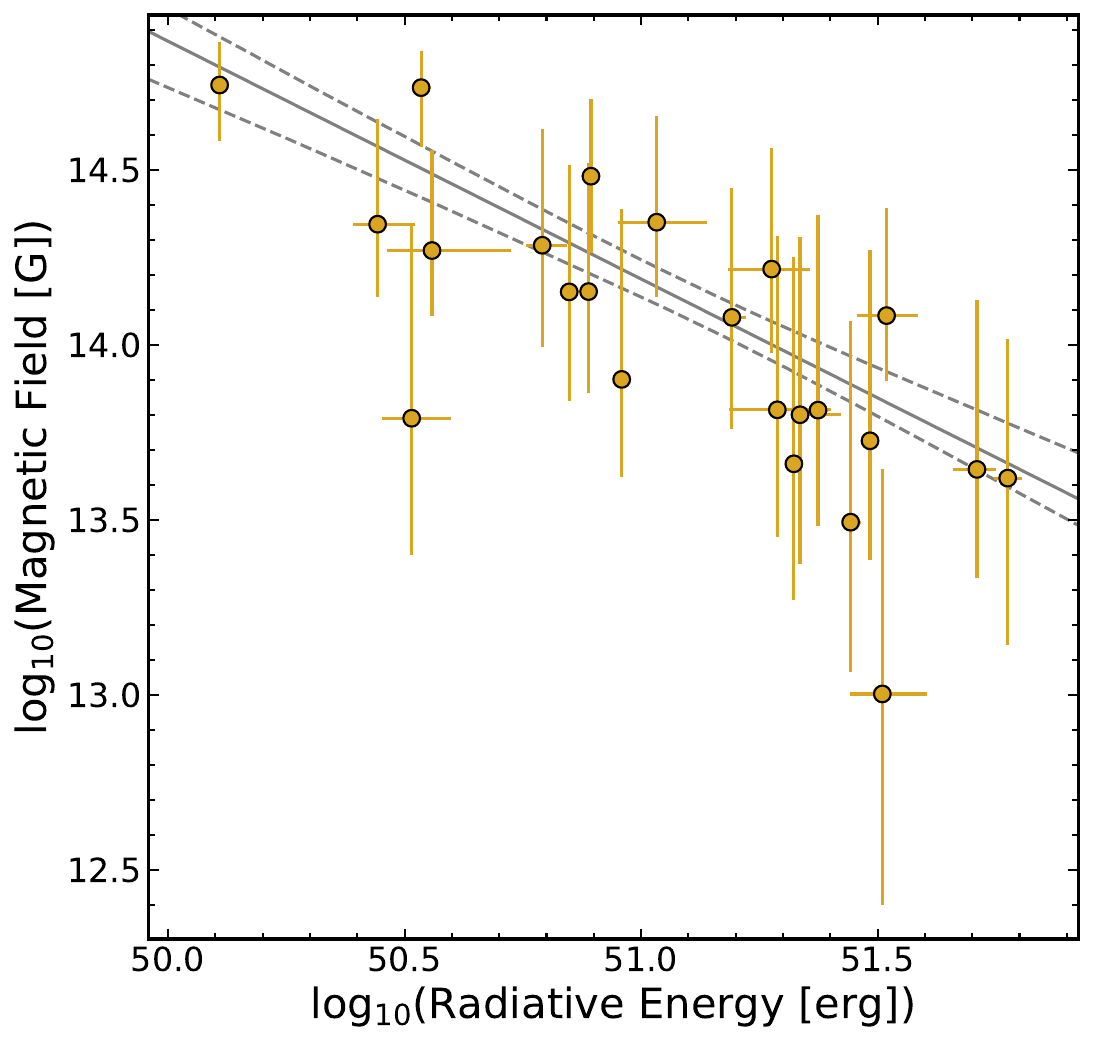}
 \caption{\textit{Upper panels}: spin period as compared to the peak UV/optical luminosity (left) and the radiative energy (right). \textit{Lower panels}:  kinetic energy (left) and magnetic field (right) as compared to the radiative energy. In all four panels, the solid gray line is the line of best fit, and the dashed gray lines are plus/minus one sigma from the best-fit line.}
 \label{fig:EL_corrs}
\end{figure*}

In addition to the comparisons of key physical parameters shown in Figure \ref{fig:MOSFiT}, we searched for correlations between these physical parameters and the peak luminosities and radiative energies, defined as the integral of the UV/optical luminosity over all times, for the SLSNe in our sample. We tested 25 correlations between the various parameters, necessitating a stricter threshold to consider a given correlation as statistically significant. Dividing the nominal p-value of $0.05$ by 25 to account for the number of additional tests, we impose a revised p-value of $\sim 0.002$ for significance. In Figure \ref{fig:EL_corrs} we show the strongest of these correlations. These correlations are as follows. In the upper left panel, we show an anti-correlation between the spin period and peak UV/optical luminosity, with Kendall $\tau = -0.59$ and a p-value of $8.2 \times 10^{-5}$. The upper right shows the anti-correlation between the spin period and radiative energy, with Kendall $\tau = -0.54$ and a p-value of $2.9 \times 10^{-4}$. The lower left shows the correlation between the kinetic energy of the ejecta and radiative energy, with Kendall $\tau = 0.54$ and a p-value of $1.7 \times 10^{-4}$. Finally, the lower right shows the anti-correlation between the magnetar magnetic field and radiative energy, with Kendall $\tau = -0.60$ and a p-value of $2.5 \times 10^{-5}$. We confirmed that these correlations exist at significant levels, whether we use the medians or best fits (using the score returned by MOSFiT) as our physical parameters. For consistency with previous MOSFiT results, we will continue to use the median values for each physical parameter.

The anti-correlations between the magnetar spin period and both the peak luminosity and radiative energy are not surprising when considering the assumptions and expectations of a magnetar central engine. The input energy from a magnetar spin-down model scales most strongly with the spin period of the magnetar, with $E_{mag} \propto P^{-2}$ \citep[e.g.,][]{ostriker71, kasen10, omand24}. While this extra energy from the magnetar must then be diffused through the ejecta \citep{arnett82}, it is clear that shorter spin periods provide larger reservoirs of additional energy to power the supernova. Conversely, a longer spin period provides less additional energy to the supernova, placing a limit on the increased radiation seen for SLSNe as compared to typical Type Ic SNe. Interestingly, in both the peak luminosity and radiative energy, there are hints of increased scatter at short spin periods, although the small sample size precludes any definitive conclusions. If astrophysical, this may indicate that below some critical spin period that there is enough additional energy from the magnetar to power an SLSN, but variations in the other physical parameters, which can change the diffusion timescale and therefore the radiative luminosity, may result in different peak luminosities and energies. 

The physical origin of the correlation between the kinetic energy of the ejecta and the radiative energy is not as straightforward under traditional magnetar spin-down models. The best-fitting linear trend is given by

\begin{equation}
\begin{split}
    \log_{10}\left(\frac{E_\textrm{kin}}{\textrm{erg}}\right) = (0.66^{+0.16}_{-0.19})\left[\log_{10}\left(\frac{E_\textrm{rad}}{\textrm{erg}}\right) - 51.1\right] \\
    + 51.6^{+0.1}_{-0.1}
\end{split}
\end{equation}

indicating a correlation at a level of $\approx3.5\sigma$. To confirm that such a correlation was not simply the result of modeling assumptions, we conducted a Monte Carlo simulation. We randomly drew parameters for 100 SLSNe assuming a Gaussian distribution centered on the median values for each parameter from \citet{nicholl17c} and a standard deviation for each parameter based on the 1$\sigma$ uncertainties from the joint posteriors of the \citet{nicholl17c} sample. We then calculated the luminosity as a function of time and the resulting integrated energy. We then applied the cut on kinetic energy discussed in Section 3.8 of \citet{nicholl17c}. This typically yielded $\approx 30$ objects, close to our sample size. We computed the Kendall $\tau$ correlation strength and significance for the set of simulated SLSNe. We repeated the whole procedure 5000 times and asked for how many realizations were the kinetic energy and radiative energy correlated as strongly (i.e., higher $\tau$) and as significantly (i.e., lower p-value) than our observed correlation. We found that only 3 out of 5000 (0.06\%) of the trials met these requirements, suggesting that this correlation is not a simple covariance introduced by the assumptions inherent to the MOSFiT modeling.

One naive explanation for the correlation between kinetic energy and radiative energy is simply that sources with high ejecta masses also have high nickel masses, providing additional energy. Under the assumption that $^{56}$Ni decay provides the energy for these SLSNe, we can use the scaling between nickel mass and energy production \citep[e.g.,][]{nadyozhin94} to estimate the fraction of the ejecta mass that must be in $^{56}$Ni. To explain all of the emitted energy through nickel decay, half of our sample requires $^{56}$Ni masses larger than the ejecta mass, which is clearly unphysical. Even for a more conservative assumption that 10\% of the emitted energy is a result of $^{56}$Ni decay requires that half of the sample has nickel masses larger than 1 M$_{\odot}$, considerably larger than other stripped-envelope supernovae \citep[e.g.,][]{afsariardchi21}. It therefore seems unlikely that this correlation between kinetic energy and radiative energy results from unmodeled $^{56}$Ni decay. 

Recent models of magnetar-powered SLSNe, which account for ejecta acceleration from interaction with the pulsar wind nebula, may more naturally explain this correlation. This is because the kinetic energy of the ejecta is boosted as the ultra-relativistic pulsar wind collides with the supernova ejecta \citep[e.g.,][]{omand24}. Thus, the kinetic energy (through the ejecta acceleration due to the pulsar wind) and the radiative energy (from the magnetar spin-down luminosity) are both driven by the magnetar spin-down and should be correlated. Indeed, the kinetic and radiative energy are each correlated with the magnetar spin period, supporting this picture.

The strong anti-correlation between magnetic field and radiative energy is likely the result of the need to match the magnetar spin-down timescale and diffusion timescale in the ejecta for an SLSNe to be luminous \citep[e.g.,][also see Fig. \ref{fig:MOSFiT}]{nicholl17c}. This may simply lead to an observational bias, where SNe that do not lie on such a correlation are not luminous and therefore are not observed at a given distance. Thus, the true distribution of these parameters in nature may not result in a strong anti-correlation. This correlation supports previous work suggesting that the magnetar spin-down and diffusion timescales must be well-matched for an SLSN to occur \citep{kasen10, metzger15, nicholl15, nicholl17c}.

\section{Conclusions} \label{sec:conclusions}

In this work, we study the UV/optical evolution of 27 well-observed SLSNe. We select only sources that have been well observed by the \swift UVOT, allowing for strong constraints on their UV emission and temperature evolution. Due to a calibration update affecting the UV bands of \swift UVOT by up to 0.3 mag in some cases, we homogeneously recompute the \swift UVOT photometry for each SLSN. The majority of our SLSNe also have long-term optical light curves enabled by modern all-sky transient surveys. Through our analysis of the SLSNe light curves, we have recovered several known trends among SLSNe. The first is that the SEDs of SLSNe are well-fit by modified blackbodies. Through a comparison of modified and simple blackbody models, we found that while many sources are fit well by either SED model, a majority of sources prefer a modified blackbody. These findings are in agreement with direct studies of SLSNe rest-frame UV spectra \citep{chomiuk11, yan17, yan18}. From our modified blackbody fits, we find a median temperature of $\approx 11,000$ K and a median radius of $\approx 4 \times 10^{15}$ cm, each consistent with previous work \citep{lunnan18, chen23a}.

While modified blackbody fits to the UVOT data provide strong constraints on temperature and radius evolution, the incomplete coverage of many events precludes the measurement of a peak luminosity and/or radiative energy for some objects. We therefore used MOSFiT and extrapolation of the best-fit model to find a median peak luminosity of $1.5 \times 10^{44}$ erg s$^{-1}$ and median radiative energy of $2 \times 10^{51}$ erg, again consistent with earlier work \citep{nicholl17c, lunnan18, angus19, chen23a}. 

With the same MOSFiT runs, we estimated key physical parameters of the SLSNe, including the neutron star spin period and magnetic field strength, ejecta mass, and kinetic energy of the ejecta. The distributions of these parameters for our sample are in full agreement with MOSFiT fits to other samples of SLSNe \citep[e.g.,][]{nicholl17c, blanchard20, hsu21, chen23b}. We find that despite correcting UV data taken when the \swift UVOT calibrations overestimated the source magnitudes, the key physical parameters remain consistent within $\sim 10$\%. One interesting trend apparent from our MOSFiT runs is a possible anti-correlation between the ejecta mass and magnetic field strength. Such a correlation is not seen in previous works \citep{nicholl17c} and may simply be a result of observational bias, as SLSNe with lower ejecta masses and weaker magnetic fields are less luminous. However, this possible anti-correlation, combined with the known anti-correlation between ejecta mass and spin period, may have implications for the formation of neutron stars during core-collapse supernovae.

We find additional correlations between physical parameters and the peak luminosity and radiative energy output of the SLSNe. The anti-correlations between spin period and luminosity and spin period and energy are caused by the spin period being the dominant factor in setting the available extra energy for the SLSNe under a magnetar model. The correlation between kinetic energy and radiative energy is inconsistent with being simply the result of additional nickel mass within the ejecta, and qualitatively agrees with models that account for ejecta acceleration from interaction with the nascent pulsar wind nebula. Finally, the anti-correlation between magnetic field strength and energy seems most related to the requirement that the diffusion and magnetar spin-down timescales are well-matched to power an SLSNe, as compared to a typical Type Ib/c supernova.

We note that our study only considers a magnetar model \citep{ostriker71, kasen10, inserra13, nicholl17c} when fitting the observed multi-band light curves of these SLSNe. A non-negligible fraction of SLSNe-I show signs of bumps or undulations in their light curves \citep{nicholl16, lunnan20, hosseinzadeh22, west23}. This behavior is not typical of a basic magnetar spin-down model, although some recent work has attempted to extend magnetar models to describe bumpy light curves \citep{chugai22, moriya22, dong23}. Furthermore, some studies have suggested that $\sim 25$\% of SLSNe-I, particularly those with light curve undulations, can be better described with a H-poor CSM interaction model \citep{chen23b}. The origin of these light curve undulations remains unclear. Despite a sample of SLSNe with well-measured UV evolution, we find no strong trends between any of the physical parameters studied here and the presence of light curve undulations. Additionally, some events like SN2018ibb, while observationally classified as SLSNe-I, may have long-term light curves that disfavor magnetar models \citep{schulze24}.

As Type I SLSNe can be very luminous, they allow for studies of supernova physics and rates at high redshift \citep{angus19}. Additionally, there is promising evidence that SLSNe-I can be used as cosmological probes \citep{inserra21, khetan23}. As such, it is important to understand the progenitors and explosion physics of such events. With the upcoming Legacy Survey of Space and Time \citep[LSST; ][]{ivezic08} on the Vera Rubin Observatory, we will find many more SLSNe. As we have shown, as long as there is sufficient rest-frame UV coverage, such events can be well-studied and used to further understand this rare population of massive star supernovae. Finally, we have provided the uniformly reduced, updated Swift photometry for these 27 well-observed SLSNe.

\acknowledgments

We thank the referee for helpful comments and suggestions that have improved the quality of this manuscript. We thank Matt Nicholl for helpful feedback on the manuscript. We also thank Steve Schulze and Conor Omand for useful comments.

J.T.H. and this work were supported by NASA award 80NSSC21K0136. B.J.S. is supported by NSF grants AST-1908952, AST-1920392, AST-1911074, and NASA award 80NSSC19K1717. 

This research has made use of the SVO Filter Profile Service (http://svo2.cab.inta-csic.es/theory/fps/) supported from the Spanish MINECO through grant AYA2017-84089. \\

\noindent \textit{Facilities:} \swift(UVOT) \citep{roming05} \\
\textit{Software:} MOSFiT \citep{nicholl17c, guillochon18}, emcee \citep{foreman-mackey13}

\bibliography{bibliography}{}
\bibliographystyle{aasjournal}

\end{document}